\documentclass[11pt]{article}

\usepackage{amssymb,amsmath}                                           
\usepackage{bbm}                                                       
\usepackage[usenames,dvipsnames]{color}                                
\usepackage{setspace}                                                  
\usepackage{hyperref}                                                  
\usepackage[left= 1in,right=1in,top=1in,bottom=1in]{geometry}          
\usepackage{comment}                                                   
\usepackage{authblk}                                                   
\usepackage{graphicx}                                                  
\usepackage{caption}                                                    
\usepackage{subcaption}
\usepackage{enumerate}													
\usepackage{epigraph}
\usepackage[utf8]{inputenc}
\usepackage[english]{babel}
\usepackage{apacite} 
\usepackage{natbib}  

\let\oldmarginpar\marginpar
\renewcommand\marginpar[1]{\oldmarginpar{\color{red}\raggedright\scriptsize #1}}
\newcommand{\pb}[2]{\ensuremath{\lf\{#1,#2 \rt\}}}

\newcommand{\diby}[2]{\ensuremath{\frac{\partial #1}{\partial #2}}}

\def\lf {\ensuremath{\left}}											
\def\rt {\ensuremath{\right}}
\def\de {{\rm d}}														

\def\L { \ensuremath{\mathcal {L}} }
\def\H { \ensuremath{\mathcal {H}} }
\def\mA { \ensuremath{\mathcal {A}} }
\def\mS { \ensuremath{D} }
\def\Lie { \ensuremath{\mathfrak {L}} }

\title{ When scale is surplus }
\date{}

\author[1,2]{{\bf Sean Gryb}\thanks{email: \href{mailto:sean.gry@gmail.com}{sean.gryb@gmail.com}, website: \href{https://seangryb.wordpress.com}{seangryb.wordpress.com}, ORCID id: 0000-0002-0439-3548}}
\author[3]{{\bf David Sloan}\thanks{email: \href{mailto:d.sloan@lancaster.ac.uk}{d.sloan@lancaster.ac.uk}}}
\affil[1]{{\it Faculty of Philosophy}, University of Groningen}
\affil[2]{ {\it Van Swinderen Institute for Particle Physics and Gravity}, University of Groningen }
\affil[3]{ {\it Physics Department}, University of Lancaster }
\affil[1]{{\it Department of Philosophy}, University of Bristol}
\affil[2]{{\it H. H. Wills Physics Laboratory}, University of Bristol}

\begin{document}

\maketitle

\begin{abstract}
	We study a long-recognised but under-appreciated symmetry called \emph{dynamical similarity} and illustrate its relevance to many important conceptual problems in fundamental physics. Dynamical similarities are general transformations of a system where the unit of Hamilton's principal function is rescaled, and therefore represent a kind of dynamical scaling symmetry with formal properties that differ from many standard symmetries. To study this symmetry, we develop a general framework for symmetries that distinguishes the observable and surplus structures of a theory by using the minimal freely specifiable initial data for the theory that is necessary to achieve empirical adequacy. This framework is then applied to well-studied examples including Galilean invariance and the symmetries of the Kepler problem. We find that our framework gives a precise dynamical criterion for identifying the observables of those systems, and that those observables agree with epistemic expectations. We then apply our framework to dynamical similarity. First we give a general definition of dynamical similarity. Then we show, with the help of some previous results, how the dynamics of our observables leads to singularity resolution and the emergence of an arrow of time in cosmology.
\end{abstract}

\clearpage

\tableofcontents
\clearpage

\epigraph{Suppose that in one night all the dimensions of the universe became a thousand times larger. The world will remain \emph{similar} to itself, if we give the word \emph{similitude} the meaning it has in the third book of Euclid. Only, what was formerly a metre long will now measure a kilometre, and what was a millimetre long will become a metre. The bed in which I went to sleep and my body itself will have grown in the same proportion. When I awake in the morning what will be my feeling in face of such an astonishing transformation? Well, I shall not notice anything at all. The most exact measures will be incapable of revealing anything of this tremendous change, since the yard-measures I shall use will have varied in exactly the same proportions as the objects I shall attempt to measure. [Original italics]}{\emph{The relativity of space} from \emph{Science and method}\\Henri Poincar\'e, 1908 (translation \citeyear{poincare2003science}). }

\section{Introduction} 
\label{sec:introduction}

\subsection{Similarity and conceptual problems in modern physics} 
\label{sub:similarity_and_conceptual_problems_in_modern_physics}

In Henri Poincar\'e's magistral \emph{Science and Method} \citep{poincare2003science} we're invited to image a universe like our own that is identical in every respect except that it is a thousand times larger. Such a universe, he argues, would be indiscernible from our own because any reference standard that could be used to measure the length of a body would have grown in exact proportion the length of that body, and therefore any measurement using this standard would be unaffected. He calls such a transformation a \emph{similarity} in reference to the same geometric symmetry discussed as early as Euclid.

Poincar\'e's argument for empirical indiscernibility under similarity can be applied more generally to measurements of temporal duration. If duration is measured using relative changes within a system --- say by recording the number of oscillations of the pendulum of a clock --- then, by Poincar\'e's argument, the transformed system should be indiscernible from the original, and therefore no change in duration can be measured.

But compared to length, representations of duration are made considerably more subtle by dynamical considerations. Useful standards of duration are defined by their ability to bring the dynamical laws into a particular form. For example, absolute time in Newtonian mechanics is defined in terms of inertial motion, which in turn is determined by the form of the dynamical laws. The precise way that a similarity transformation acts on the representations of a dynamical system is therefore affected by how the temporal standards depend on the laws. This requires a more comprehensive approach.

The most important consideration in defining such an approach is the way in which the form of the laws affects the transformation properties of the velocities. In Lagrangian theories, length and time standards are related by the conventions used to define the unit of Hamilton's principal function. Heuristically speaking, this unit defines a standard of angular momentum that can be used to convert lengths to velocities using some convention for inertial mass. We will call similarity transformations that take into account the dynamical considerations involved in defining velocity \emph{dynamical similarities} because of their relation to the transformations of the same name first studied in the context of the Kepler problem (see Section~\ref{sub:DS as subsystem} for the explicit connection).

The primary goal of this paper is to give a detailed analysis of dynamical similarity. We will show that, while often recognised in different guises, dynamical similarity is an under-appreciated symmetry in many physical systems with significant implications for major conceptual problems in modern physics. To achieve this goal, we need a philosophical framework for understanding symmetry that is general enough to handle the peculiar features of dynamical similarity. In developing such a framework, we accomplish several noteworthy secondary goals. In particular, we give an account of symmetry that identifies surplus structure\footnote{ In this paper it will be convenient to define surplus structure as eliminable structure according to one of the senses of surplus structure discussed in \cite{nguyen2020surplus,bradley2020representational}. This is because, in the examples we will discuss, dynamically similar surplus structure will be explicitly eliminable. It is, however, possible to rephrase our discussion of dynamical similarity by taking surplus structure to be a representational redundancy. This will be shown in future work. } using the freely specifiable initial data of a theory required for empirical adequacy. We then use this account to resolve certain tensions between epistemic and dynamic definitions of symmetry in various well-studied cases including the Galileo ship and the Kepler problem.

We will then apply our framework to a cosmological setting and find that dynamical similarity relates surplus structure. We will then review previous technical results\footnote{ Mostly notably those developed in \cite{Koslowski:2016hds} and \cite{Sloan:2019wrz}. }  that show how many kinds of singularities in general relativity --- including those of certain important big bang and black hole solutions --- are removed when dynamically-similar structure is taken to be surplus. Finally, we will see that a novel explanation for the arrow of time arises for such systems.

Central to our arguments is our analysis of symmetry. We will develop a principle, which we will call the \emph{Principle of Essential and Sufficient Autonomy (PESA)}, that we will use to identify the observable structures of a theory. While the PESA can be applied to general symmetries, it will be introduced here in order to accommodate certain non-standard formal properties of dynamical similarity that prevent an analysis in terms of standard techniques. The PESA gives a clear prescription for formally characterising the models that a particular theory has hypothesised to be empirically identical.\footnote{ We use the word `model' in the philosopher's sense as the representation of a possibility. This usually translates, in physics jargon, to an `history' or sometimes a `solution' of a theory. } If such a theory is found to be empirically adequate, then the PESA suggests a strong norm for classifying as surplus any structures that vary between identified models and as observable any structures that remain unchanged. Such an identification of models can also be motived by the Principle of the Identity of Indiscernibles (PII) since the relevant models agree with respect to all empirically accessible properties. The two-step application of hypothesising observable structure using the PESA and then checking for empirical adequacy sidesteps tensions between dynamic and epistemic definitions of symmetry discussed at length in the literature.\footnote{ See for example \cite{vanFraassen:sym_and_surplus,roberts2008puzzle,healey2009perfect,belot:sym_and_equiv,dasgupta2016symmetry,pittphilsci16622,belot2018fifty} and the more extended discussions in Section~\ref{sub:dynamics_versus_epistemology}. } Our proposal separates the formal properties of symmetry (using the PESA) from contingent facts about the world (by checking for empirical adequacy). More speficially, the PESA gives precise dynamical criteria that can be used to distinguish observable and surplus structure by considering the number of independent input data required to solve the evolution equations. In this way our novel proposal represents a distinct alternative to category-theoretic (cf. \cite{weatherall2016understanding,nguyen2020surplus,bradley2020representational}) definitions of surplus structure or other recent proposals for analysing symmetry (cf. \cite{CAULTON2015153,dewar2019sophistication,read_martins_sophistry}) that do not provide specific dynamical criteria that can be applied to dynamical similarity. To test our formalism, we will apply it to some well-studied physical examples\footnote{ These cover many of the prominent situations discussed in \cite{kosso2000empirical,brading2004gauge,greaves2014empirical,belot:sym_and_equiv,pittphilsci16623} and \cite{saunders2013rethinking,knox2014newtonian}. See Section~\ref{sec:assessing_indiscernibility_examples} for details. }. We reproduce orthodox expectations regarding the epistemic considerations of the symmetries of those systems. Passing these tests give us confidence that our proposal can be reliably applied to the novel cosmological applications of dynamical similarity considerer in this paper.

After establishing the reliability of our framework we will turn attention to dynamical similarity. Following and expanding upon \cite{Sloan:2018lim}, we will give a general definition of dynamical similarity, illustrate how it is importantly different from other symmetries in physics, and develop a general scheme for eliminating surplus structure due to dynamical similarity in different contexts. In applying the PESA in light of the empirical adequacy of different physical theories, we will find that when dynamical similarities act on approximately (but not completely) isolated subsystems of the universe, the removal of dynamically similar structure is \emph{not} justified. This applies to many well-studied cases in physics including the so-called Runge--Lenz symmetries of the Kepler problem (see \cite{belot:sym_and_equiv}). Here, our analysis is consistent with standard interpretations. But when dynamical similarities act on the universe, our formalism prescribes that the removal of dynamically similar structure \emph{is} justified. This applies, in particular, to modern theories of cosmology. 

To be more concrete about the implications of removing dynamically similar structure, we will develop a conceptual framework, leveraged on previous formal results, in which pathologies in the evolution equations of certain cosmological systems will be seen to be due to the surplus structure introduced by dynamical similarity. These pathologies can then be removed if the symmetry is understood as relating indiscernible states of the universe. This leads to a form of singularity resolution in a classical theory of gravity. Moreover, mathematical features unique to dynamical similarity will be shown to define certain privileged structures --- namely \emph{attractors} and so called \emph{Janus points} (to be defined in Section~\ref{sub:friction_janus_points_and_attractors}) --- on the space of indiscernible models of cosmological theories. In addition, the dynamics on the space of discernible states will be seen to have a fundamentally \emph{frictional} character. The privileged structures and the friction-like behaviour establish a preferred arrow of time for observers that find themselves in a state near an attractor. This suggests a novel solution to the problem of the arrow of time similar to that proposed in \cite{Barbour_2014} and \cite{barbour:janus}. Finally, the central role played by the standard of angular momentum in Lagrangian theories with dynamical similarity suggests exciting implications for the foundations of statistical and quantum mechanics. Removing the ``extra mathematical hooks'' of dynamical similarity thus has significant implications for many empirical and conceptual problems in modern physics.

\subsection{The dynamics of similarity} 
\label{sub:dyn of sim}

In the previous section we described how a similarity transformation should act on the representations of a dynamical system. We highlighted the subtle role of the dynamical laws in determining the transformation properties of the velocities. We will now make these comments more precise by giving a formal definition of dynamical similarity. This definition will provide the starting point for the analysis of the remainder of the paper.

Our definition must involve a statement of the laws themselves because these define the convenient temporal standards for the theory. It will thus be helpful in our analysis to restrict to a general class of dynamical systems called \emph{Lagrangian systems}. A Lagrangian system is a system whose dynamically possible models (DPMs), $\gamma_\text{DMP}$, are those curves on the state space of the theory that are stationary points of an action functional, $S[\gamma]$, such that $\delta S[\gamma]|_{\gamma_\text{DPM}} = 0$. Given such a system, we define a \emph{dynamical similarity} as a transformation that rescales the action functional by a constant:\footnote{ More specifically, dynamical similarities will be maps on certain representational structures (see Section~\ref{sub:general_definition}) of the theory that induce a pullback of $S$ that is a rescaling of it. }
\begin{equation}\label{eq:DS gen}
	S[\gamma] \to c S[\gamma']\,.
\end{equation}
Under this definition, if $\gamma$ is a DPM then $\gamma'$ is also a DPM because \eqref{eq:DS gen} preserves the stationary condition of $S[\gamma]$. Dynamical similarities, when they exist, therefore map DPMs to DPMs, and consequently preserve the condition for satisfaction of the laws.

To understand how such transformations match the intuitions of the previous section, consider that the units of $S$ are those of angular momentum, and therefore that the effect of a dynamical similarity is to actively rescale the global standard of angular momentum for the system.\footnote{ Throughout this paper, we will use an \emph{active} notion of dynamical similarity whereby the dynamical similarities will be thought of as rescaling the standard of angular momentum of the system. } Different choices of $c$ therefore represent different ratios between the standard of length and the standard of velocity. Using these transformations, one can then determine how to rescale the standards of length and time to produce a similarity transformation that preserves the condition for satisfaction of the laws. In Section~\ref{sub:general_definition}, we will see more explicitly how to do this in general (e.g., Equation~\ref{eq:local S}) and in Sections~\ref{eq:kepler universal} we treat two examples that help to build intuition for dynamical similarity and the situations in which it should be expected to have important implications for physics.

We end these introductory remarks with two important comments that will figure prominently in our discussions below.

First, dynamical similarities are somewhat unconventional symmetries in that they do not preserve the formal relationships between position and momentum.\footnote{ I.e., dynamical similarities do not preserve symplectic structure. These features will be expanded upon in Section~\ref{sub:a_non_symplectic_symmetry}.} This distinguishes them from other well-known gauge symmetries of physics such as those of electromagnetism, the standard model of particle physics and general relativity. Because of these differences, dynamical similarities cannot be treated using standard physics tools that assume symmetries of a standard form.\footnote{ E.g., Noether's theorems and Dirac's algorithm. The PESA is compared with Dirac's algorithm in Section~\ref{sub:the_pesa}. } The PESA, however, will apply to dynamical similarity by construction. As we will see, these peculiarities are what leads to so many of the interesting and noteworthy features of dynamical similarity.

Second, as we will see in Section~\ref{sub:DS as subsystem}, well-known applications of dynamical similarity concern subsystems of the universe and take dynamical similarities to transform between \emph{discernible} models of a theory. As an example, dynamical similarities transform between distinct planetary orbits consistent with Kepler's third law. It is therefore important that our proposal reproduce these intuitions while also justifying our novel treatment of cosmology. We will see in Sections~\ref{sub:DS as subsystem} and \ref{sub:friedmann_lema^itre_robertson_walker_cosmology} that this is indeed the case.

Third, standard practice in cosmology recognises that \emph{similarity} transformations relate indiscernible states of the world. It is an uncontroversial fact that the observational data in cosmology are independent of the convention used for fixing spatial size.\footnote{ To be more precise, the observational data is independent of the initial value of the scalar factor. } What is not generally recognised, however, are the \emph{dynamical} implications of this fact. In particular, the surplus structure associated with \emph{dynamical} similarity is usually retained within the formalism. The remarkable consequences for cosmology that we will explore in this paper are a direct consequence of explicitly treating dynamical similarity as one would any symmetry that relates indiscernible states of the world.

\subsection{Prospectus} 
\label{sub:prospectus}

In this paper we develop a general framework for symmetries that distinguishes the observable and surplus structures of a theory by using the minimal freely specifiable initial data for the theory that is necessary to achieve empirical adequacy. This framework is then applied to well-studied examples including Galilean invariance and the symmetries of the Kepler problem. We find that our framework gives a precise dynamical criterion for identifying the observables of those systems, and that those observables agree with epistemic expectations in terms of a standard universal--subsystem distinction. We then apply our framework to dynamical similarity. First we give a general definition of dynamical similarity. Then we show, with the help of some previous results, how the dynamics of our observables leads to singularity resolution and the emergence of an arrow of time in cosmology.

Our general procedure will be developed in Section~\ref{sec:PESA}. First, we will set up a framework for symmetries in Section~\ref{sub:a_framework_for_symmetry} that distinguishes two kinds of symmetry: \emph{universal symmetries} that act on the universe as a whole and \emph{subsystem symmetries} that act on isolated subsystems. Knowing which context --- universal or subsystem --- applies to any given theory is a subtle matter. We will propose the PESA, which we define in Section~\ref{sub:the_pesa}, to provide clear criteria for identifying the surplus structure of a theory, and therefore clearly distinguishing between these contexts.

We will then test our proposal in Section~\ref{sec:assessing_indiscernibility_examples}. First we explore the well-studied Galilean boosts as examples of universal (Section~\ref{sec:universal Gal}) and subsystem (Section~\ref{sec:Gal ship}) symmetries. In Section~\ref{sub:DS as subsystem} we then explore the Kepler problem, and see that Kepler's third law is consistent with dynamical similarity being a subsystem symmetry. In all these examples, the PESA is seen to reproduce orthodox views. In Section~\ref{eq:kepler universal} we develop intuition for dynamical similarity as a universal symmetry by considering hypothetical models related to the Kepler problem.

The remaining sections of the paper will be devoted to giving a more precise definition of dynamical similarity, which expands upon the formalism presented in \cite{Sloan:2018lim}, and giving some interesting applications to cosmology. We will highlight the main representational features of dynamical similarity in Section~\ref{sub:general_definition}, and outline a general procedure for writing a dynamically-similar description of Lagrangian systems in Section~\ref{sub:geometric_structure}. In Section~\ref{sub:friction_janus_points_and_attractors} we will then describe how some of the interesting features resulting from dynamical similarity arise.

Finally, in Section~\ref{sec:applications} we review some results of \cite{Koslowski:2016hds} and \cite{Sloan:2019wrz} and show how our framework can be used to address several important conceptual problems in modern physics. In Section~\ref{sub:cosmo probs}, we show how removing the surplus structure due to dynamical similarity can resolve the initial singularity (i.e., the \emph{big bang}) in an important class of cosmological theories. And in Section~\ref{sub:sing res in GR} we describe how these results may generalise to a larger class of singularities in general relativity. As a final application in Section~\ref{sub:aot} we discuss how dynamical similarity leads to a new proposal from solving the problem of the arrow of time. In Section~\ref{sec:conclusions} we conclude by summarising and discussing some potential applications to the foundations of thermodynamics and quantum mechanics.

\section{The Principle of Essential and Sufficient Autonomy} 
\label{sec:PESA}

\epigraph{Methods for removing excess structure are much more than mopping up procedures. They are not something merely to be done \emph{after} our representations have been crafted, like portraitists erasing stray pencil marks, or sculptors removing extra clay. Methods for removing excess structure are the very heart of theorizing.}{ \cite[Ch. 23, p.390]{vanFraassen:sym_and_surplus}. }

\subsection{The very heart of theorizing} 
\label{sub:a_framework_for_symmetry}

In this section we will introduce a powerful normative principle, central to our analysis, that we will call the \emph{Principle of Essential and Sufficient Autonomy (PESA)}. This principle identifies the representational structures of a theory that are observable, and distinguishes these from those that are surplus. It can thus be used to identify and remove excess structure. The PESA is closely related to what Barbour has called \emph{Poincar\'e's principle}\footnote{ The principle was introduced in \cite{Barbour_Bertotti}. See Section~5 of \cite{barbour2010definition} for the historical connection with Poincar\'e. } but differs importantly in the way that it evades certain ontological subtleties that we will outline below.

As we will see, the virtues of the PESA are not necessary obvious when contemplating a theory purely at the formal level. Rather, the PESA's appeal is most apparent during the process of theory-construction. Within this process, the PESA is an extremely useful conceptual and formal tool for developing and refining a particular theory.

In \cite{vanFraassen:sym_and_surplus}, theory-construction is described as a two-stage process:
\begin{quote}
	There are two stages in theory construction. The first is to generate a set of models rich enough to embed the phenomena, the second is to attempt to simplify those models by exposing and eliminating excess structure. Continuing in this way the structure of the models is pared down, being careful not to jeopardize their capacity to embed the phenomena. [p.390]
\end{quote}
We find this characterisation compelling because it highlights the delicate balance that must be achieved between wanting to reduce the amount of surplus structure and retaining the ability to describe the phenomena. The PESA aims to find this balance by imposing precise mathematical criteria. But before we can state the principle explicitly, it will first be necessary to address some known problems related to defining symmetry.

\subsubsection{Dynamics versus epistemology} 
\label{sub:dynamics_versus_epistemology}


The broad topic of `symmetry' has been extensively discussed in the philosophical literature. An important question in these discussions is what, if any, metaphysical or epistemological conclusions can be drawn from symmetries. Perhaps unsurprisingly, the answers provided to this question depend upon the way that symmetries are defined. Defining symmetry involves some well-known difficulties that originate from a tension between the dynamical and epistemic aspects of symmetry. This tension can be used to divide definitions of symmetry into those that make use of the dynamical properties of a theory or those that make use of epistemic considerations.\footnote{ An alternative way to characterise the difference between `dynamic' and `epistemic' definitions of symmetry is in terms of `theoretical' versus `empirical' symmetries as defined in \cite{healey2009perfect}. }

\emph{Dynamic} definitions of symmetry define symmetries as structure-preserving transformations between a theory's DPMs. Such definitions have proved useful, for example, when the dynamic symmetries are compared with the space-time symmetries of a theory according to the analysis of \cite[\S3.4]{earman1989world}. In such approaches the formal invariances of the underlying mathematical structures of a theory are believed to support reliable inferences about the ontological properties of the world.

\emph{Epistemic} definitions of symmetry define symmetries as transformations between a theory's indiscernible models. In such definitions, it is an epistemic consideration --- namely, indiscernibility --- that is used as the primary criterion for defining a symmetry transformation. Such attempts are exemplified by the approach of \cite{dasgupta2016symmetry} linking good notions of symmetry to undetectability.

While some authors (e.g., \cite{pittphilsci16622}) have framed the tension between dynamic and epistemic definitions as a dialectic between opposing camps, it is easy to find authors who find difficulties with both views. These difficulties arise through the existence of examples where dynamic and epistemic considerations lead to incompatible conclusions. Using such examples, \citet[\S1]{belot:sym_and_equiv} shows how the ``unhappy'' combination of dynamic and epistemic considerations leads to a ``recipe for disaster'' when trying to provide a precise definition of symmetry. Similar examples can be found in \cite{roberts2008puzzle}, where the tension is framed as a ``puzzle.''

A vivid way to illustrate the origin of this tension is to consider systems where symmetries are seen to have \emph{direct empirical consequences}. Examples where this occurs are given in \cite{kosso2000empirical} and further details have been discussed in \cite{brading2004gauge,greaves2014empirical,friederich2015symmetry,pittphilsci16981}. In such cases, symmetry transformations that act uniformly on the representations of a subsystem can fit the dynamical definition of a symmetry --- in that they preserve the condition for satisfaction of the laws --- but not the epistemic definition --- in that they relate empirically distinguishable states.

Perhaps the most well-studied example is that of \emph{Galileo's ship}, which will be discussed in detail in Section~\ref{sec:Gal ship}. In this example, a ship is moving almost frictionlessly with respect to a shore. Galileo boost transformations of the ship that leave the shore fixed preserve the condition of satisfaction of the laws but relate \emph{discernible} states of the system. We will have much more to say about symmetries of this kind later. But for now we note that they are exemplars for illustrating how dynamic definitions of symmetry do not always agree with purely epistemic definitions.

To sidestep difficulties of this kind, we will attempt to avoid metaphysical considerations as much as possible in our analysis of symmetry. Instead, we will focus on the ability, or inability, of a theory to describe phenomena. We will try to draw a clear separation between the formal definition of a theory and the contingent fact of whether that theory can usefully describe the world. By insisting on this separation, we can refrain from making ontological commitments based on the purely formal properties of a theory that will put dynamical definitions of symmetry in conflict with epistemic considerations.

Because of this, we will not be able to draw strong metaphysical conclusions from fromal definitions of symmetry alone. But this does not mean that nothing useful can be learned from our analysis of symmetry. Quite to the contrary, we will be able to give precise dynamical criteria, in terms of the amount of independently specifiable initial data of a theory, that will exactly indicate the empirical structure that a particular theory hypothesises. Moreover, these criteria will be used to naturally distinguish different theoretical contexts where the formal structures of a theory bear different hypothesised relations to the phenomena. This leads to a rich view of symmetry that can be usefully applied to a multitude of different physical systems. 

\subsubsection{Kinds of symmetry} 
\label{sub:contexts_of_symmetry}

Given the considerations of the previous section, we will find it useful, on the one hand, to define symmetries in a more `dynamic' tradition in terms of formal transformations that preserve certain representational structures of a theory. At a minimum, we will require that a symmetry preserve a theory's DPMs. However, symmetries can, and usually do, preserve additional structures. When they do we can characterise symmetries according to those structures (e.g., `space-time symmetries' preserve certain space-time structure).\footnote{ A framework for analysing symmetry in this way for the purposes of understanding temporal symmetry in quantum gravity will be presented in \cite{time_regained}. This is in the same spirit as category-theoretic definitions of symmetry in terms of structure-preserving isomorphisms such as those discussed in \cite{weatherall2016understanding,nguyen2020surplus,bradley2020representational}. }

We will also find it useful, on the other hand, to appeal to a more `epistemic' tradition by distinguishing different theoretical contexts that characterise the relationship between the formally defined symmetries of a theory and the empirical properties of the world. In this way, we will allow for symmetries to be transformations both between surplus structure and certain measurable quantities. Guided by the examples discussed above where symmetries are seen to have direct empirical consequences, we will define two kinds of symmetry that will distinguish different theoretical contexts where symmetries will relate both discernible and indiscernible states.  

To make our language more precise, we will first gives some definitions. We will call \emph{appearances} structures in the world that can be described by experimental and measurement reports.\footnote{ Although our language roughly follows \cite[p.64]{van1980scientific}, the distinction we wish to draw between `surplus' and `observable' structure should not be confused with the more subtle and complex observable-unobservable distinction.}. We then call \emph{observable structure}, or simply \emph{observables}, the (suitably interpreted) representational substructures of the models of a theory that are candidates for describing observable phenomena. In particular, the observables of a theory are those structures that a theory hypothesises can be mapped bijectively to appearances. Conversely, we will call \emph{surplus structure} any representational structures in the theory that cannot be mapped bijectively to appearances. Finally, we say that a theory is \emph{empirically adequate} if the theory has one model whose observable structures are isomorphic to all appearances.

We pause briefly to clarify a potential ambiguity in our terminology. We have decided to adopt the term `observable' to refer to the representational structures that a particular theory hypothesises to be isomorphic to the appearances when the theory is empirically adequate. This is importantly \emph{not} the sense of `observable' used by Van~Fraassen to denote structures discernible by the senses. Instead, it is the sense of `observable' used by practising physicists when describing the gauge-invariant representational structures of a theory. As we will see in detail in Section~\ref{sub:the_pesa} below, the notion of observable we have in mind will be determined explicitly by the form of the laws of the theory by the way in which those laws are well-posed. As such, our usage is perhaps best understood as a generalisation of the notion of \emph{Dirac observable}, which is made in terms of canonical constraints implied by the action principle of the theory. This point is emphasised in Section~1.5.2 of the standard text of \cite{henneaux1992quantization}:\footnote{ Note that they refer to \emph{Dirac observables} as \emph{classical observables}. }
\begin{quote}
	Indeed, no information other than the action principle was required in the determination and classification of the constraints. Thus, the action itself enables one to decide what are the observables.
\end{quote}
Our definition is in this spirit but does not rely on canonical constraints and, as we will see, will be applicable to more general contexts such as that of dynamical similarity.

With this usage in mind, we can use our language to define a \emph{universal symmetry} as a symmetry that leaves invariant all the observable structures of a theory. Canonical examples of universal symmetries are the bulk diffeomorphism symmetries in general relativity and gauge symmetries of classical electromagnetism and Yang--Mills theory.\footnote{ When gauge-conserving boundary conditions have been imposed (cf. \cite{belot2018fifty}). } It is important to emphasise that this definition is independent of whether a theory with a universal symmetry is empirically adequate or not. We can then define a \emph{universal context} in which the empirical adequacy of a theory with a universal symmetry is being investigated. It is only once this context has been specified that different prescriptive ontological commitments, pertaining for example to surplus structure, can be assessed.

We define a \emph{subsystem symmetry} as a symmetry that leaves invariant some, but not all, observable structures of a theory.\footnote{ This definition matches the formal definition of a \emph{regional symmetry} as defined in \cite{pittphilsci16981}, but where the structures that shift the region are observable in the theory in question. We defer to that paper for a more technical definition. } The subsystem in question is simply the set of observable structures on which the symmetry acts non-trivially. A canonical example of a subsystem symmetry is the boost symmetry of a Galileo ship. We can then define a \emph{subsystem context} in which the empirical adequacy of a theory with a subsystem symmetry is being investigated. We note that the terms `universal' and `subsystem' do not reference the domain of applicability of the theory itself. Thus, universal symmetries can arise in theories that are themselves not considered to be universal (e.g., the $U(1)$ invariance of electromagnetism).

\subsubsection{Two problems for a theory} 
\label{sub:two_deficiencies_of_a_theory}


In this section, we will try to motivate the PESA by highlighting two important ways in which a theory can be problematic in being able to describe phenomena. The PESA will then be designed to identify these deficiencies. This identification will be a subtle task in the cosmological examples we will consider later. To begin, we must first be clear about what we will understand a theory to be.

We will understand a theory to broadly constitute two things: i) a statement of its DPMs by a specification of its basic mathematical structures and laws, and ii) a collections of hypotheses about how these models are related to structures in the world. These hypotheses include assumptions about measurement procedures, and must provide maps between observables and appearances.\footnote{ In the \emph{DEKI} account of representation developed in \cite{frigg2020modelling}, these maps correspond to a \emph{key} for how interpreted representations (i.e., our `observables') are used to exemplify properties that can be imputed onto the target system. } In this way a theory, on our view, must itself specify all of its observables. Given this way of understanding a theory, one \emph{first} hypothesises about which quantities are observables, and \emph{then} checks for empirically adequacy as a test of these hypotheses. Because of this, whether a particular symmetry is universal or subsystem could depend simply on which structures the theory identifies as observable.

There are two ways that a theory's observables can be inadequate in describing the phenomena of the world. In the first case, a theory can have too \emph{few} observables to describe all the appearances. Such a theory cannot be empirically adequate and must either be rejected or extended. This situation is usually straightforward to diagnose because there will be phenomena that the theory can't describe.

In the second case, a theory can have too \emph{many} observables than are strictly necessary for describing the appearances. Here, if the system is deterministic then the dynamics of the observables are underdetermined by the phenomena. The evolution equations of the theory therefore cannot be solved in terms of the observables in question.

It might seem that such underdetermination would be obvious to diagnose. But because it is often possible to fix the extraneous observables by making various auxiliary assumptions (often implicitly), this second case is usually more subtle to diagnose than the first. In this case, a theory may be believed to be empirically adequate --- i.e., it may be believed to have an isomorphism between observables and appearances --- even though it posits unnecessary observables, and therefore the map between observables and appearances is in fact many-to-one. Such a misdiagnosis can lead to faulty conceptualising that misinterprets surplus for observable structure. We will see that this is precisely what happens with dynamically-similar structure in the applications of Section~\ref{sec:applications}. In those cases, the PESA will be our primary guide towards deciding the appropriate set of observables for describing the phenomena.

\subsection{The PESA} 
\label{sub:the_pesa}

We have now compiled the necessary ingredients to state the PESA. Consider a theory whose representations involve a general state $\psi(t,b^I(t))$ at time $t$ that is completely specified by the time-dependent structures $b^I(t)$, which form a generating set for some algebra $\mathcal B$. Consider then that the time evolution of the state is determined from the $b^I(t)$ according to equations of the form
\begin{equation}\label{eq:red ev}
	\dot b^I(t) = f^I(b^I, t)\,,
\end{equation}
for some functions $f^I$ for which we won't specify any significant restrictions. This form of the laws is very general as any higher order system can be re-written as a first order system by adding structures to the set $B = \{b^I\}$. In general, however, the equations \eqref{eq:red ev} will not be well-posed in terms of initial data that can be determined by the appearances alone. By `well-posed' we mean that a specification of arbitrary $b^i(t_0)$ at some time $t_0$ uniquely determines the solution $b^i(t)$ at some later time $t$ for all $t>t_0$ relevant to the theory. The PESA makes use of this fact to give a dynamic definition of a theory's observables that can be used to identify surplus structure.

Let $\mathcal P$ denote the space of all possible distinct appearances at a time $t$ hypothesised by a theory. The PESA states that the \emph{observable algebra} $\mathcal A$ of the system should be generated by the subset $A = \{a^i\} \subseteq B$ such that the restriction of \eqref{eq:red ev} to $A$, i.e.,
\begin{equation}\label{eq:gen ev}
	\dot a^i(t) = \tilde f^i(a^i, t)\,,
\end{equation}
is well posed and that, for the maps $\mathcal O$ of the form $\mathcal O: \mathcal A \to \mathcal P$,\footnote{ The representational structures $a^i$ are assumed to be equipped with a suitable interpretation. When this is done, the maps $\mathcal O$ can be roughly thought of as providing a \emph{key} as defined in the DEKI account of representation developed in \cite{frigg2020modelling}. }
\begin{enumerate}[(i)]
	\item The equations \eqref{eq:gen ev} are \emph{autonomous} in $A$.\label{crit:autonomy}
	\item The maps $\mathcal O:\mathcal A \to \mathcal P$ are onto; i.e., the algebra $\mathcal A$ has \emph{sufficient} structure for representing $\mathcal P$.\label{crit:sufficiency}
	\item The maps $\mathcal O$ are one-to-one; i.e., the algebra $\mathcal A$ has a \emph{necessary} (or \emph{essential}) amount structure for representing $\mathcal P$.\label{crit:necessity}
\end{enumerate}
Observables are then defined to be elements of $\mathcal A$. Note that the PESA consists of two non-trivial steps. First, requiring that the evolution of the system be reducible to \emph{well-posed} equations of the form \eqref{eq:gen ev}. And second, requiring that the criteria (i)-(iii) are satisfied for these evolution equations. Focus on both of these steps is what leads to the novelty of the PESA.

The criterion \eqref{crit:autonomy} implies that the time $t$ be uniquely specifiable in terms of a generating set $A$ of the observable algebra, and that the states of a theory can be specified without referencing an external time parameter. The criterion \eqref{crit:sufficiency} requires that the appearances do not underdetermine the initial data required to solve the evolution equations while the criterion \eqref{crit:necessity} ensures that the appearances do not overdetermine these initial data.

The autonomy requirement is motivated by the observation that the PESA has prescriptive force only when the dynamical laws of the theory in question depend on a closed set of structures. If the laws were to depend on additional external structures, then the PESA would be incapable of assessing whether these external structures are observable or not. The autonomy requirement therefore demands that one consider all structures --- particularly the evolution parameter(s) --- involved in solving the dynamical equations when determining the observable algebra of a theory.\footnote{ The autonomy requirement is also very helpful for treating dynamical similarities since these usually rescale the evolution parameter, and so the invariant system is usually expressed as an autonomous set of differential equations.}

The two criteria of sufficiency and necessity are directly motivated by the two ways discussed in the previous section that a theory can be inadequate in accounting for the phenomena: namely that the theory can either have too few or too many observables. These criteria are also clearly motivated by the two stages of theory-construction highlighted in the quote from \cite{vanFraassen:sym_and_surplus} at the beginning of Section~\ref{sub:a_framework_for_symmetry}. The PESA is simply a way to formalise these intuitions into precise mathematical criteria that link the well-posedness of the evolution equations of a theory to its observables. In practice, the value of the PESA is in emphasising that one should explicitly test for each criterion --- autonomy, sufficiency and necessity --- separately.

To clarify how the principle should be used in practice, under the PESA \emph{all} structures that must be specified in order to solve a computational algorithm for determining the future state of a system should be part of the generating set of the observable algebra. This includes initial or boundary data for the dynamical equations. But it also includes \emph{all} other parameters of the theory such as particle masses and couplings constants. In this way, certain structures can easily be under-counted when constructing the observable algebra.

Conversely, the generating set of the observable algebra can also be over-counted because it is not always obvious that a structure can be removed without compromising the ability of a theory to define the maps $\mathcal O$. For instance, in a theory where gauge-fixing conditions are used to give a unique evolution to the equations of a system, the map between gauge-fixed structures and appearances is not one-to-one because many gauge-fixing conditions can be used to describe the same phenomena. The PESA would therefore prescribe against identifying gauge-fixed and observable structures because of the necessity criterion~\eqref{crit:necessity}. We will see important examples in Section~\ref{sub:sing res in GR} where certain structures can be removed from the observable algebra of general relativity without affecting the predictions for cosmology. In such cases, the PESA prescribes that the theory should be expressible in terms of the smallest possible autonomous algebra.

The PESA can be usefully compared and contrasted with standard techniques for analysing surplus structure in the physics literature. A standard technique that is similar to the PESA, called the \emph{Dirac algorithm} after the procedure developed in \cite{dirac2001lectures} (see \cite[\S1.5]{henneaux1992quantization} for a modern treatment), also defines an observable algebra using the formal properties of the evolution equations. The observable algebra in question defines the \emph{Dirac observables} as representational structures in terms of which the evolution equations are well-posed.\footnote{ More specifically, the Dirac observables are those functions on the phase space of a Hamiltonian representation of a theory that commute with all the first class constraints of the theory.\label{ftn:Dirac obs} } The Dirac observables, however, are defined using specific formal structures that exist only for certain kinds of theories (see footnote~\ref{ftn:Dirac obs}). Moreover, even in cases where the Dirac observables are formally defined, there is disagreement about their correct physical interpretation when it comes to time evolution in parametrization-invariant systems.\footnote{ See critical discussions in \cite{barbour2008constraints,pons2005dirac,pons2010observables,kuchar1991problem,kuchavr2011time,pitts2014change,gryb2016schrodinger}. } As we will see in Section~\ref{sub:a_non_symplectic_symmetry}, theories where dynamically similar structure is surplus cannot be analysed using such techniques because the formal properties of dynamical similarity violate the conditions under which the Dirac algorithm can be applied. The PESA (or something like it) is therefore required for an analysis of dynamical similarity and may also be useful in understanding the interpretation of Dirac observables with regards time reparametrization invariance. 

It should be emphasised the necessity criterion~\eqref{crit:necessity} is not simply an Occamist norm. If a theory identifies observables that are unnecessary, then that theory may provide false explanations or predictions that rely on the behaviour of the unnecessary observables. We note, however, that the PESA aims makes no metaphysical claims about the explanatory role of surplus and observable structures. Instead it highlights the fact that these play different formal roles in finding well-posed evolution equations. Moreover, the PESA does not demand an \emph{explicit} characterisation of the observable algebra. In many theories, such as general relativity and Yang--Mills theory, no such explicit construction is available. In such cases an \emph{implicit} characterisation of the observable algebra, using for example dynamical constraints, that formally identifies the maps $\mathcal O$ is sufficient. Our formalism is thus compatible with the different notions of surplus structure discussed in \cite{dewar2019sophistication,nguyen2020surplus,bradley2020representational}.

In the universal context, the autonomy requirement removes any dependence of the evolution of a theory's state on temporal structures that do not exist within the theory. In the subsystem context, if there is a temporal structure external to the subsystem in question, then this structure must exist somewhere in the observable algebra of the theory.

\subsubsection{Poincar\'e observables} 
\label{sub:poincare_observables}

The observable algebra selected by the PESA is similar to a proposal made at least as early as Poincar\'e, and which has been re-emphasised in \cite{Barbour_Bertotti} in the form of \emph{Poincar\'e's principle}. One of the key ways in which the PESA differs from these earlier proposals is that, in separating a theory's formal symmetries from contingent facts, the PESA is agnostic to metaphysical claims about the world. The principle is therefore neither committed to Poincar\'e's instrumentalism nor to Barbour's relationalism. Importantly, the PESA is \emph{not} committed to a strong relational ontology even though it also takes inspiration from Poincar\'e's writings. Since Barbour has linked Poincar\'e's principle to his form of relationalism, we find it useful to take a moment to illustrate how Poincar\'e's writings are compatible with the PESA independently of relational assumptions.

In Chapter 7 of \emph{Science and Hypothesis} \citep{poincare_foundation}, Poincar\'e studies the case of relative motion and in particular is considering the ontological status of the absolute orientation of a Keplerian system in absolute space. He is particularly concerned with the role of angular momentum effects in such a system if it is treated as isolated, where ``thick clouds hide the stars from men who cannot observe them, and even are ignorant of their existence.''[p.109] In this context, he asks: ``How will those men know that the earth turns round?'' In our language, he is considering the role of rotational symmetry in the universal context.

He notices that the interpretation of angular momentum in this system is ambiguous and can be treated in at least two ways: on the one hand as a constant of motion (which he calls an `accidental constant') in an experimentally undetectable absolute space, and on the other hand as a constant of nature (which he calls an `essential constant') encoded in the relational laws of the epistemically accessible universe. In the end, he concludes that choosing between the two amounts ``only the choice of hypotheses.'' He emphasises that what matters are the precise amount of data required to describe the phenomena: [p.114]
\begin{quote}
	``Provided that the future indications of our instruments can only depend on the indications which they have given us, or that they might have formerly given us [\textit{autonomy}], such is all we want [\textit{sufficiency}], and with these conditions we may rest satisfied [\textit{necessity}].'' [Italicised words added for emphasis.]
\end{quote}
In the quote above, we've indicated how the criteria of autonomy, sufficiency and necessity, which are required by the PESA, appear to play an essential role in Poincar\'e's analysis. Moreover, he focuses on the total number of independent data required to find the future state of the system, which is how we defined the observable algebra above. For these reasons, we will refer to the elements of the observable algebra specified by the PESA simply as \emph{Poincar\'e observables} in analogy to the \emph{Dirac observables} commonly defined in the physics literature on gauge theories.

\subsection{The PESA and empirical adequacy} 
\label{sub:the_pesa_and_indiscernibility}

It is clear that Poincar\'e had strong empiricist intuitions in mind when emphasising the role of what we have called Poincar\'e observables. While he admits that the conventional interpretation of universal angular momentum in terms of changes of orientation in absolute space is ``the most convenient solution for the geometer'', he adds that ``it is not the most satisfactory for the philosopher, because this orientation \emph{does not exist}'' [emphasis added].

While we will not find it necessary to endorse Poincar\'e's metaphysics, we do think that something can be usefully gained by considering the Poincar\'e observables. When a theory is found to be empirically adequate, we will take the view that \emph{any transformations that leave the Poincar\'e observables invariant relate states that are empirically identified.} When a theory is empirically adequate, there is an isomorphism between the Poincar\'e observables of one of a theory's models and the appearances. Thus, only changes in the Poincar\'e observables can lead to empirically discernible states. States where the Poincar\'e observables take the same value are therefore identical in all their observable properties. In this way, our view is also partially motivated by the Principle of the Identity of Indiscernibles (PII).

We can compare the PESA when applied to an empirically adequate theory to standard treatments of symmetry given in the philosophical literature. We restrict our discussion to a proposal made in \cite{CAULTON2015153} because of its strong similarities to our own proposal. Caulton begins by defining what he calls an \emph{analytic} symmetry. In our terminology, an analytic symmetry corresponds roughly to a universal symmetry as we have defined it.\footnote{ Some notable differences: there is no requirement in \cite{CAULTON2015153} that a symmetry relate DPMs and structures are assumed to take specific values (i.e., they are treated as \emph{quantities}). } The proposal is then described in two phases:
\begin{quote}
	During the first phase we set up representational links between the theory and the observable portion of the physical world, under the assumption that the theory is empirically adequate (or similar). In the second phase, we maximise the theory’s analytic symmetries, taking advantage of the representational links forged in the first phase so as not to compromise empirical adequacy. [\S 4]
\end{quote}
This procedure is clearly very close to the one we endorse in this paper. Once the appropriate representational links have been set up, maximising a theory's analytic symmetries is analogous to the necessity criterion~\eqref{crit:necessity}. The sufficiency criterion~\eqref{crit:sufficiency} is automatically satisfied when the theory is empirically adequate. Thus, the main difference between the proposals is our emphasis on having an observable algebra in terms of which the dynamical system can be expressed as a set of well-posed and autonomous evolution equations. More specifically, the PESA links observable structure to those structures that provide the necessary and sufficient input data for the evolution equations. Such a requirement does not exist within Caulton's proposal. It is inspired by the Dirac algorithm but applicable to more general transformations such as dynamical similarity.

The two-step process of specifying observables using the PESA and then checking for empirical adequacy will be our main methodology for analysing dynamical similarity. Using examples we will now see that this process is consistent with the standard treatment of gauge symmetries in physics.


\section{Assessing (in)discernibility in familiar examples} 
\label{sec:assessing_indiscernibility_examples}

\subsection{Galilean boosts} 
\label{sub:galilean_boosts}

We will now illustrate the power of the PESA by applying it to some well-studied examples of symmetry in the literature and show that it eliminates tensions between dynamic and epistemic considerations. The first example we will consider is the that of the Galilean boost symmetries. These act on a Newtonian system by transforming the velocities, $\dot{\vec q}_I$, of rigid bodies (or point particles) by a constant vector $\vec a$:
\begin{equation}\label{eq:Gal boosts}
	\dot{\vec q}_I\to \dot{\vec q}_I + \vec a\,,
\end{equation}
where $I$ ranges over the total number of bodies in the system. The Lagrangian for the theory is
\begin{equation}\label{eq:Gal L}
	\L = \sum_I \frac{m_I}2 \lf(  \dot{\vec q}_I \rt)^2 - V(\vec q_I)\,,
\end{equation}
where $m_I$ are the masses of the bodies and $V(\vec q_I)$ is the potential energy function invariant under Euclidean transformations of the centre-of-mass positions $\vec q_I$ of the bodies.

\subsubsection{Barbour--Berttoti theory: boosts as a universal symmetry}\label{sec:universal Gal}

To understand the sense in which the Galilean boost symmetries \eqref{eq:Gal boosts} can be thought of as universal symmetries of Newtonian mechanics, we first recall some well-studied facts about Newtonian mechanics. Newton was able to produce an empirically adequate theory of planetary motion by assuming that isolated systems are described by indiscernible states related by Galilean boosts acting on the system. This fact is stated and proved in Corollary~V of Newton's \emph{Principia} \citep{newton1999principia}.\footnote{ Corollary~VI states that this is also the case for isolated systems undergoing \emph{arbitrary} linear accelerations. }

An efficient way to understand this invariance in a modern language is to follow a procedure developed in \cite{Barbour_Bertotti} where an isolated Newtonian system is actively boosted by introducing auxiliary fields $\vec b(t)$ that are allowed to be arbitrary functions of time. The boosted Lagrangian involves a transformation of the velocities of the form $\dot{\vec{q}} \to \dot{\vec{q}} + \vec b(t)$ and is therefore given by
\begin{equation}
	\L_\text{BB} = \sum_I \frac{m_I}2 \lf(  \dot{\vec q}_I + \vec b \rt)^2 - V(\vec q_I)\,.
\end{equation}
The Euler--Lagrange equations for $\vec b$ say that the system must be invariant under arbitrary infinitesimal boosts. These equations imply the vanishing of the total linear momentum of the system:
\begin{equation}\label{eq:BB cond}
	\vec P_\text{tot} = \sum_I \vec p_I = 0\,,
\end{equation}
where $\vec p_I = \diby{\L_\text{BB}}{\dot{\vec q}^I}$ are the generalised momenta of the system.

This condition can be satisfied by going to a reduced description where the centre-of-mass position (and velocity) has been removed from the system. Such a description can be obtained, for instance, by defining the centre-of-mass variables $\vec Q_I = \vec q_I - \vec q_\text{cm}$, where $\vec q_\text{cm} = \sum_I m_I \vec q_I$, and then eliminating one such variable in terms of the others (the choice is arbitrary). In such a reduced description, the variable $\vec q_\text{cm}$ no longer appears. This indicates that, in BB theory, changing the centre-of-mass position (and velocity), even in a time dependent way, has no effect on the variables satisfying the dynamical constraints \eqref{eq:BB cond}. Importantly, satisfaction of these constraints \emph{reduces} the number of independently specifiable initial data for the system by $6$: the centre-of-mass position and velocity, which are underdetermined by the equations of motion of the BB system. The largest, well-posed autonomous system that satisfies \eqref{eq:BB cond} is therefore $6$ dimensions smaller than the original Newtonian system.

Moreover, when the dynamical constraints \eqref{eq:BB cond} are satisfied, it is easy to see that the original Lagrangian transforms in such a way that
\begin{equation}\label{eq:symp Gal}
	S \to S + \phi\,,
\end{equation}
for some constant $\phi$, and therefore that Galilean boosts preserve the stationary points of the action. They are thus a symmetry of the theory by our definition.

The PESA can now be used to identify observables and provide clarity on the status of global boosts in Newtonian mechanics. For isolated systems, the remarkable success of Newtonian mechanics illustrates that coordinates $\vec Q^I$ (with one eliminated) can be used to write an empirically adequate theory of planetary motion. Moreover, Corollary V, or equivalently the analysis of Barbour and Berttoti outlined above, shows that the necessary and sufficient autonomous system required to predict future states from initial states does \emph{not} require a specification of the centre-of-mass velocity of the system, which is arbitrary. In other words, the smaller, well-posed posed system satisfying \eqref{eq:BB cond} is nevertheless sufficient for describing the appearances of Newtonian mechanics. Thus, the PESA prescribes that the centre-of-mass velocity is \emph{not} a Poincar\'e observable because it is not necessary \eqref{crit:necessity}.\footnote{ Autonomy can be achieved in all the examples studied in Section~\ref{sec:assessing_indiscernibility_examples} by expressing the equations of motion using \emph{Jacobi's} principle following the analysis of \cite{Barbour_Bertotti}. Doing this does not change the quantitative predictions of any of these theories. } Moreover, because global Galilean boosts leave the Poincar\'e observables invariant, they are universal symmetries of Newtonian mechanics by our definition. This agrees with modern orthodoxy regarding the role of global boosts in Newtonian mechanics as expressed for instance in \cite{saunders2013rethinking} and \cite{knox2014newtonian}.

\subsubsection{Galileo's ship: boosts as a subsystem symmetry}\label{sec:Gal ship}

In the previous section, we focused attention on the case where the boosts act on isolated systems. We now turn attention to the more familiar context in which the boosts act uniformly on approximately isolated subsystems of the universe but leave the rest of the system's structures invariant.

This context is nicely illustrated by the \emph{Galileo ship} scenario. In this scenario, a ship is floating frictionlessly (or nearly so) on water in such a way that it has uniform velocity relative to a shore. Galileo famously remarked that if you are below deck and all the windows of the ship are closed, then it would not be possible to detect observables differences between situations where the ship was moving with different uniform velocities relative to the shore \citep[pp. 186-187 (Second Day)]{galileo_dialogues}:
\begin{quote}
	``...have the ship proceed with any speed you like, so long as the motion is uniform and not fluctuating this way and that. You will discover not the least change in all the effects named, nor could you tell from any of them whether the ship was moving or standing still.''
\end{quote}
The `effects named' in the above quote are examples given by Galileo --- such as the flight path of butterflies or the trajectories of fish swimming in a bowl. These phenomena can be represented as rigid bodies in the Newtonian system described by the Lagrangian \eqref{eq:Gal L}. The ship and all the bodies in it form a subsystem of the universe that behaves, for all practical purposes, as if it is isolated from the rest of the system.

What is different from this context and that treated in the previous section is that here there are means to detect relative motion between the ship and the shore. As Galileo observers, if one climbs above the deck of the ship ``more or less noticeable differences would be seen in some of the effects noted.'' In this case, empirical adequacy cannot be achieved if the centre-of-mass velocity of the ship is allowed to vary arbitrarily: changing the velocity of the ship will obviously lead to discernible changes in the ship's motion relative to the shore.

Unlike in the universal case, empirical adequacy can only be achieved if the relative velocity of the ship and shore is taken to be a Poincar\'e observable. No BB-like constraint of the form \eqref{eq:BB cond} can be used to reduce the system to a smaller, well-posed system that is empirically adequate. This implies that the Galilean boosts of Galileo's ship are subsystem symmetries by our definition: they act invariantly on the ship and the shore but not invariantly on the relative velocity of the ship and shore. In this case, the the PESA prescribes that states states where the ship is boosted relative to the shore should \emph{not} be empirically identical. In other words, boosts of this kind should have empirical consequences. The dynamical criteria of the PESA and the empirical adequacy of Newtonian mechanics (in the energy regime in question) therefore reproduce the standard expectations.

\subsection{Kepler's third law} 
\label{sub:kepler_s_third_law}

\subsubsection{Dynamical similarity as a subsystem symmetry}
\label{sub:DS as subsystem}

In the previous section we showed how the PESA can be used in conjunction with empirical adequacy to reproduce orthodox views regarding the status of boost symmetries in Newtonian mechanics. In this section, we will showcase the power of the PESA by applying it the less familiar symmetry of dynamical similarity. As we will see in Section~\ref{sub:a_non_symplectic_symmetry}, dynamical similarities have formal features that make them inapplicable to standard symmetry treatments such as Noether's theorems or the Dirac constraint algorithm.\footnote{ Although there are generalisations of Noether's first theorem that apply to contact systems \cite{bravetti2021geometric}, and therefore may be applicable to the case of dynamical similarity. } But because the PESA applies to any theory whose laws can be expressed as dynamical equations, it can be applied to any general system that has a dynamical similarity. We begin our analysis with the well-studied context of dynamical similarity arising in the Kepler problem to illustrate how the PESA can be applied more generally. We will show explicitly at the end of this section that the dynamical similarities that act in the Kepler problem exactly fit our definition of a subsystem symmetry.\footnote{ Incidentally, this makes dynamical similarity one of the oldest known dynamical symmetries.} It is helpful to note that the dynamical similarity transformations discussed here are simply the large group transformations of the infinitesimal \emph{Runge--Lenz} symmetries described cryptically in \cite{belot:sym_and_equiv} and \cite{pittphilsci16623} (to see this compare the transformations of this section with equations~(62) of \cite{prince1981lie}).

To see the relationship between dynamical similarity and Kepler's third law, consider two bodies co-orbiting under an inverse square law of attraction. It is well-known (see for example \cite[\S3.7]{goldstein2000classical}) that the trajectories of these bodies are described by conic sections, which in turn are defined by their semi-major axis and eccentricity. We denote the separation of the bodies by $r$. In the context we are interested in this section, which is also the context in which Kepler's laws where originally proposed, the two-body system is further described by the angle $\theta$ made by a line between them to a fixed orientation with reference to some idealized distant stars. For simplicity we can absorb all constants such as the masses and the gravitational coupling into the single constant $C$ such that the Lagrangian for the system takes the form:
\begin{equation}
\L= T-V = \frac{1}{2} \left(\frac{\de r}{\de t}\right)^2+ \frac{r^2}{2}  \left(\frac{\de \theta}{\de t}\right)^2 + \frac{C}{r} \,.
\label{eq:KeplerLagrangian}
\end{equation}
We will now define a transformation $D$ that acts on both $r$ and $t$ while leaving $\theta$ and $C$ fixed in such a way that the Lagrangian is rescaled. Let us pick a positive real number $\beta$. If we let $D: r\rightarrow \beta r$ and $D: t\rightarrow \beta^\alpha t$ we see that if we choose $\alpha=3/2$ then the Lagrangian is rescaled under $D: \L \rightarrow \beta^{-1} \L$. Because $D$ rescales $\L$ and therefore the action, it is a dynamical similarity by our definition. This in turn implies that $D$ preserves the stationary points of the theory, and is therefore a symmetry.

An immediate consequence of $D$ being a symmetry of the Kepler problem is consistency with Kepler's third law, which can be stated as follows:
\begin{quote}
	The square of a planet's orbital period is proportional to the cube of the length of the semi-major axis of its orbit.
\end{quote}
In particular, this means that for every closed solution to the two body problem with a semi-major axis length $l$ and period $T$, there exists a solution with semi-major axis  $\beta l$ whose orbital period is $\beta^{3/2} T$ for all positive real numbers $\beta$. This follows trivially from the action of $D$ because of the scaling properties of length, via the transformation of $r$, and duration, via the transformation properties of $t$.

Because $D$ acts equally on points along a trajectory, both semi-major and semi-minor axes are equally rescaled, and hence any solution is mapped onto one that has the same eccentricity.\footnote{Note that Kepler's third law applies also to solutions with difference eccentricities and is, therefore, a stronger claim that what can be proved with dynamical similarity alone.} Under this transformation the total energy of the system, $H=T+V$, transforms non-trivially but surfaces of constant energy are merely rescaled: $D:H \rightarrow \beta^{-1}H$, and therefore their shape is preserved. Likewise the angular momentum is also rescaled: $J=r^2 \dot{\theta} \rightarrow \sqrt{\beta}J$, as is expected from the fact that the action, and therefore the standard of angular momentum, is rescaled in the same way.

We now turn to the question of whether the Kepler problem is formulated in the universal or the subsystem context. If the two bodies are taken to represent an idealisation of a planet-sun system in our solar system, then a transformation that changes the energy and angular momentum of this two-body system is clearly intended to represent a discernible change. This is because the transformation $D$ is intended to act uniformly on one particular planet-sun system while keeping the motion of other planets and the distant stars fixed.

The PESA therefore prescribes, by the sufficiency criterion \eqref{crit:sufficiency}, that in an empirically adequate theory $\beta$ should be a Poincar\'e observable. To determine $\beta$ the value of the rescaled semi-major axis and orbital period can be compared with the value of the semi-major axes and orbital periods of the remaining planets in the solar system. Since $D$ acts invariantly on the observables of the two-body subsystem but does not leave invariant the Poincar\'e observable $\beta$, $D$ fits our definition of a subsystem symmetry of the Newtonian $N$-body problem. Kepler's laws are therefore clearly applicable to the subsystem context. As in the case of Galileo's ship, dynamical similarities in the Kepler problem can be seen to imply direct empirical consequences. Once again, the PESA and the empirical adequacy of Kepler's laws has reproduced orthodox expectations.

\subsubsection{Dynamical similarity as a universal symmetry}
\label{eq:kepler universal}

In this section we imagine a hypothetical thought experiment where dynamical similarity is being studied for the Kepler problem in the universal context. This is similar to the hypothetical world considered by Poincar\'e in Chapter 7 of \emph{Science and Hypothesis} that we described in Section~\ref{sub:poincare_observables}, but adapted to dynamical similarity rather than to rotation. In this thought experiment, we imagine that there is no other matter content in the universe aside from two self-gravitating bodies and a collection of infinitely distant stars that have fixed relative configurations amongst themselves on the celestial sphere i.e., they are literally dots on the celestial sphere. These stars provide a reference frame for determining the orientation of a line connecting the two bodies expressed as an angle, $\theta$, between this line and one of the distant stars. While $\theta$ can be empirically determined using the system of distant stars as a reference, in this world the distance $r$ between the two bodies cannot be compared to any other length.

On the other hand, the \emph{relative} instantaneous rate of change, $\frac 1 r \frac{\de r}{\de \theta}$, that uses the dimensionless angle $\theta$ as a clock can be empirically accessible because the \emph{relative} change of the size of the system can be instantaneously compared with rate of change of $\theta$. This is a possible observable quantity that can be represented in a dynamical system where instantaneous change is empirically accessible. Because there is one internal clock and two dimensionless velocities in terms of this clock, we expect an autonomous system of three variables to be required for empirical adequacy in this hypothetical world.

Our task now is to find a representation of this autonomous system. We start with a representation of the system in terms of the configuration variables $r$ and $\theta$ and their velocities as measured by some auxiliary clock $t$. We take the Lagrangian for the system to be that of the standard Kepler Lagrangian \eqref{eq:KeplerLagrangian}. The empirical inaccessibility of $r$ suggests that it is surplus, and that it should be removed. The subtlety, as always, lies in removing the dependence of $t$ on the length standard used to define $r$. Because this standard determines the unit of the action, we simply need to find a set of variables for the entire isolated system that is invariant under the dynamical similarity transformation $D$ defined in the previous section. In other words, we are looking for a representation of the system where dynamical similarity is treated as a universal symmetry.

A general procedure for reducing an isolated dynamical system to an autonomous system invariant under dynamical similarity was developed in \cite{Sloan:2018lim}. An outline of this procedure is given in Section~\ref{sub:geometric_structure} with a brief technical description in Footnote~\ref{ftn:geo contact}. The output of the procedure is a particular (non-unique) representation of an algebra of invariants for the system and a projection of the dynamical equations generated by the Lagrangian onto this space of invariants. For the purposes of this section, we will simply give one such representation. In Appendix~\ref{sec:homo V} we show how the representation used in this section can be generalised to an arbitrary homogeneous potential.

It is straightforward to explicitly verify that the base elements $a=\sqrt{r} \dot{r} = \frac{2}{3} \frac{\de}{\de t} \left(r^{3/2}\right) $, $b=r^{3/2} \dot{\theta}$ and $\theta$ form an algebra invariant under $D$.\footnote{In this representation of dynamical similarity the constant $C$ is taken to be invariant under $D$. This set of invariants can be constructed from the symplectic 2-form and the action of $D$ using $\rho = r^{1/2}$ (see Footnote~\ref{ftn:geo contact} for the definition of $\rho$ and a brief description of this construction).} We can construct this set by determining the transformation properties of the velocities and multiplying by the appropriate power of $r$ that leaves the combination invariant. In terms of this algebra, it is a short calculation to show that the evolution equations of \eqref{eq:KeplerLagrangian} reduce to the autonomous system of equations:\footnote{ The first order form of this system highlights the number of independent initial data and the conditions for integrability --- both of which are integral to the PESA. }
\begin{align}\label{eq:reduced eoms}
\frac{\de b} {\de \theta} &= \frac{a}{2} & \frac{\de a}{\de \theta} = \frac{a^2 - 2C}{2b} + b\,,
\end{align}
where $\theta$ has been used as an internal clock. This dynamical system is therefore equivalent to the original dynamical system defined by \eqref{eq:KeplerLagrangian} but with all standards of length removed (including those appearing in temporal intervals). This three-dimensional reduced system is autonomous and has the necessary and sufficient number of base elements for achieving empirical adequacy in this hypothetical world. The PESA therefore prescribes that $a$, $b$ and $\theta$ should be Poincar\'e observables for this theory. It is important to note that the system defined by \eqref{eq:reduced eoms} is not a standard Hamiltonian system (we will see why we expect this more generally in Section~\ref{sub:a_non_symplectic_symmetry}), and therefore standard tools for analysing symmetries, such as Noether's theorems and the Dirac algorithm, are not applicable. It is a distinct advantage of the PESA that it can provide a concrete specification of the observable algebra for this system.

Clearly it is difficult to have clear intuitions for what phenomena could be observed in such a hypothetical world. Adding only a bit of structure, however, allows us to see how the PESA does lead to intuitive results, and greatly illuminates how more conventional physics is embedded in the formalism of dynamical similarity in the universal context.

The difficulty in the previous example is that the orbital radius $r$ and period $T$ rely implicitly on spatial and temporal standards, and are therefore not invariant under dynamical similarity. If, however, we make use of spatial and temporal standards \emph{within} the system, then a more intuitive picture emerges.

We can add such structures to our hypothetical system by considering an idealisation of the Earth-Sun system where the Earth is taken to be a uniform sphere\footnote{ This idealisation serves only to fix the factor of proportionality between $R^2$ and the momentum of inertia. It can therefore be straightforwardly generalised.} of radius $R$ that has a constant rotational velocity $\dot\phi$ about its axis of rotation. The angle $\phi$ can be determined by marking a fixed position on the Earth and taking the angle between this and a reference star on the rotational plane. The Lagrangian for this system is:
\begin{equation}
	\L_\text{ext} = \frac{ \dot r^2}2 + \frac {r^2 \dot \theta^2} 2 + \frac{R^2 \dot \phi^2} 5 + \frac C r\,,
\end{equation}
where again we have absorbed units of mass and factors of $G$ into the constant $C$.

The equation of motion for $R$ tells us immediately that $R$ is a constant of the motion --- as expected for the radius of the Earth. We can therefore use it to set a natural standard for length in the system. By multiplying by the appropriate powers of $R$, we can easily construct an invariant notion of radial distance between the Earth-Sun system by forming the ratio: $\gamma = \frac r R$ and an invariant momentum using $p = R^{1/2} \dot r = R^{3/2}\dot\gamma$. Noting that $\theta$ and $\phi$ are already invariant, we can construct the remaining invariants by rescaling the canonical momenta appropriately by $R$:\footnote{This set of invariants can be constructed by using $\rho = R^{1/2}$. See Footnote~\ref{ftn:geo contact} for details.}
\begin{align}
	p_\theta &= \frac{r^2 \dot \theta}{R^{1/2}} = \gamma^2 R^{3/2} \dot \theta &
     p_\phi &= R^{3/2} \dot \phi\,.
\end{align}
In terms of these variables, time derivatives always appear in the equations of motion with the pre-factor $R^{3/2}$, which cancels the transformation of $t$ under $D$ by construction.

We can obtain a scale-free autonomous system by switching to an internal clock. For this system, a natural choice is a rescaling of $\phi$ such that $R^{3/2} \frac{\de \phi}{\de t_\phi} = 1$, where we define $t_\phi \propto \phi$. Such a choice is always available for this system because $\dot \phi$ and $R$ are constants of motion of the original system. Physically, the clock $t_\phi$ can be interpreted as reading out a sidereal day. In terms of $t_\phi$, a short calculation shows that the equations of motion for the system reduce to:
\begin{align}
	p &= \gamma' & p' &= - \frac C {\gamma^2} \\
	\theta' &= \gamma^2 p_\theta & p_\theta' &= 0\,, 
\end{align}
where primes indicate derivatives with respect to $t_\phi$. We thus obtain the equations of motion for the original Kepler system \eqref{eq:KeplerLagrangian} but with length standards given by the radius of the Earth and temporal standards given by the sidereal day.

This example illustrates that the PESA reproduces our intuitions for a closed dynamical system with internally defined standards for length and duration. This provides more evidence that the PESA is a reliable symmetry principle for when we apply it in a novel way to the cosmological applications of Section~\ref{sec:applications}. The dynamical similarity $D$ acts on representations of length in such a way that the privileged rod, in this case of radius of the Earth, gets rescaled in exactly the same way as the distances being measured, in this case the radial distance between the Sun and the Earth. Similarly, representations of time intervals are also rescaled so that the number of sidereal days in a sidereal year is fixed. One obtains the standard Kepler system precisely because the rods and clocks used to describe the autonomous motion are dynamically isolated from the rest of the system.

Let us make a couple important comments to help build intuition for the role of dynamical similarity in physical systems. The case treated above is an example of a more general result that a conservative system can always be approximately recovered from a dynamically similar one when the internal rods and clocks used to describe the motion are approximately dynamically decoupled from the system itself. In the Kepler system, this also provides a vivid illustration of the difference between the universal and subsystem contexts: a subsystem transformation that rescales the orbital radius and period but keeps the length $R$ and and sidereal time $t_\phi$ fixed will produce an empirically distinct model of the system. This is one way (and perhaps an historically accurate way) to understand Kepler's third law for fixed eccentricity. In the universal context one simply investigates indiscernibility under different representations of the same internal clocks and rods.

What makes the standard Kepler picture work is therefore the dynamical isolation of the internal clocks and rods. This, however, is necessarily an idealisation because in order to gain knowledge of the value of any clock or rod, an observer must interact with it --- even if only very weakly. In practice, because there is no gravitational screening, no clock or rod can ever be perfectly isolated from the system it is describing. In the Earth-Sun system, the sidereal day is affected by, among other things, the tidal drag between the Earth and the Moon, and is therefore not a dynamically isolated clock. Indeed, this is an example of a much more general discussion about the role and limitations of idealised inertial clocks dating back at least as far as Mach.\footnote{ For a list of notable discussions of these points see \cite{barbour2001discovery,barbour:mach_before_mach,mach1907science,pfister2014ludwig}. } Dynamically similarity becomes particularly important in such cases because the system can no longer be idealised by a conservative system and instead exhibits friction-like behaviour. As we will see in Section~\ref{sub:cosmo probs}, nowhere is this more important than in the cosmological case. There, the expansion of the universe affects all clocks in the universe and implies that, on cosmological time-scales, the system is non-conservative. As we will see, this leads to significant implications for several conceptual problems in cosmology.

\section{Dynamical similarity} 
\label{sec:dynamical_similarity}

In this section we give a general definition of dynamical similarity for Lagrangian systems and a brief outline of some of the geometric structures that underlie the framework. We will see that a key mathematical tool in our analysis will be the odd-dimensional cousin of symplectic geometry called \emph{contact geometry}, and that conservative Hamiltonian systems exhibiting dynamical similarity will generically have a description in terms of \emph{contact Hamiltonian systems}. For more details on the link between dynamical similarity and contact Hamiltonian mechanics, see \cite{Sloan:2018lim}. For a more details regarding the mathematical construction of contact Hamiltonian systems see Appendix~4 of \cite{arnol2013mathematical} and \cite{2019JMP....60j2902D,Bravetti_2017}. \cite{Bravetti:2018rts} gives a recent review with applications to thermodynamics and \cite{geiges2008introduction} gives a detailed list of proofs of all relevant theorems.

\subsection{General definition} 
\label{sub:general_definition}

We are primarily interested in understanding the action of dynamical similarity in Lagrangian systems in the universal context. In this context, states related by dynamical similarity are thought to be empirically indiscernible, and a mathematical procedure for identifying such states warranted. We will now give such a procedure by defining equivalence classes of DPMs under dynamical similarity. The equivalence relations thus obtained defines a projection from the original system to a smaller autonomous system that is invariant under dynamical similarity. In this way, the procedure is analogous to the reduction of a system by a gauge transformation as is familiar to standard gauge theory. We will not pursue a complete geometric construction here, but will outline a coordinate-based approach, motivated by a powerful theorem, that we will illustrate in the applications of Section~\ref{sec:applications}.

\subsubsection{A non-symplectic symmetry} 
\label{sub:a_non_symplectic_symmetry}

We begin by highlighting the way in which dynamical similarities are importantly different from most standard gauge symmetries studies in the literature. This will also allow us to give a more precise definition of dynamical similarity. Consider a general Lagrangian system described by the action functional
\begin{equation}
	S[\gamma] = \int_\gamma \de t\, \L(q, \dot q) \,,
\end{equation}
where $\L(q, \dot q)$ is the Lagrange density of the system; $(q, \dot q)$ are generalized coordinates and their velocities providing coordinates for the tangent bundle, $T\mathcal C$, over the configurations space $\mathcal C$, $t$ is the time parameter, and $\gamma$ are the kinematically possible models of the theory. As stated in the introductory Section~\ref{sub:dyn of sim}, the DPMs for this system satisfy
\begin{equation}\label{eq:eoms}
	\delta S[\gamma]\big\rvert_{\gamma_\text{DPM}} = 0\,.
\end{equation}

Now consider the general class of transformations $\mS$ that act on the basic structures of a theory\footnote{ We will be more explicit about what these structures are in Section~\ref{sub:the_action_of_dynamical_similsubarity}. } such that the pullback of the action $S$ by $\mS$ is
\begin{equation}\label{eq:gen dyn sym}
	\mS^* S \to c S + \Phi\,,
\end{equation}
for some constant $\Phi$ and some non-zero positive constant $c$. The transformation $\mS$ is the most general transformation that preserves the DPMs of the theory.

An important way to characterise the different symmetries of $\mS$ can be made using the Hamiltonian formalism. This formalism exists when one can write the Lagrangian in the form $\L \de t = p \de q - H \de t$, where the term $\theta = p \de q $ is called the \emph{symplectic potential}. The symplectic potential can be thought of as providing the relationship between the generalised momenta $p$ and the velocities $\dot q$ via the definition $p = \diby{\L}{\dot q}$, and is therefore central to the formation of geometric structures on phase space. An example of such a structure is the natural volume element on phase space called the \emph{Liouville measure}, which can be conveniently written in the language of differential forms or the coordinates $(q,p)$ as
\begin{equation}\label{eq:Liouville}
	\mu_L(R) = \int_R (\de\theta)^N = \int_R \de^N p\, \de^N q\,,
\end{equation}
where $N$ is the dimension of $\mathcal C$ and $\de$ is the exterior derivative operator on phase space. An important requirement for the Hamiltonian to exist is therefore that the volume-form defined by $\mu_L(R)$ be non-degenerate.

To understand the significance of the symplectic potential, consider what happens when the theory undergoes a transformation that shifts it by an exact differential $\theta \to \theta + \de \phi$. Under such a transformation, $\L \de t \to \L \de t + \de \phi$, where $\phi$ is an arbitrary phase space function, and the action is shifted by a constant: $S[\gamma] \to S[\gamma] + \phi\bigr\rvert_{t_0}^{t_1}$.\footnote{ Note that we therefore have $\Phi = \phi\bigr\rvert_{t_0}^{t_1}$. } This case thus corresponds to a symmetry $\mS$ in which $c = 1$. Motivated by the fact that the exterior derivative operator on phase space satisfies $\de^2 = 0$, we can define the \emph{symplectic 2-form} $\omega = \de \theta$ that characterises \emph{symplectic structure}. Symmetries for which $c = 1$ preserve the symplectic 2-form, and we will therefore refer to them as \emph{symplectic symmetries}.\footnote{ Symplectic symmetries are often referred to as \emph{divergence} symmetries (see, for example, \cite{belot:sym_and_equiv} and \cite{olver2000applications}) because they shift the Lagrangian by an exact form. }

It is difficult to overstate the importance of the symplectic structure defined by $\omega$ in formulating the Hamiltonian mechanics of dynamical systems. We've already mentioned how the natural notion of volume on phase space can be defined in these terms. Perhaps even more important is the fact that the dynamical equations themselves in the form of Hamilton's equations are also expressed in terms of the inverse of $\omega$. Even the quantum formalism is motivated by the preservation of the anti-commutation relation between $q$ and $p$ that arise through the definition of $\omega$. It is thus not surprising that symplectic symmetries play an important role in physics and are the subject of most of literature on this topic. A simple example of symplectic symmetries is the Galilean boost symmetries of Section~\ref{sub:galilean_boosts} --- as is illustrated by Equation~\ref{eq:symp Gal}. Indeed, using techniques similar to those used in that section it is straightforward to show that all the Galilean symmetries are symplectic symmetries of Newtonian mechanics.

Dynamical similarities, on the other hand, correspond to the case where $c \neq 1$ according to the definition given in Section~\ref{sub:dyn of sim}, and are therefore non-symplectic symmetries. It is rather remarkable that the physics and philosophy literature has focused almost entirely on symplectic symmetries while almost ignoring the obvious generalisation to $c \neq 1$.\footnote{ An exception being the \emph{scaling} symmetries discussed in \cite{belot:sym_and_equiv}, which can be shown to be dynamical similarities. } In fact, all the interesting features of dynamical similarity that will be introduced in Section~\ref{sub:friction_janus_points_and_attractors} and studied in the applications of Section~\ref{sec:applications} are due to the non-symplectic nature of the dynamical similarities. Moreover, it is because of the non-symplectic character of dynamical similarity that standard symmetry techniques do not apply, and why an analysis in terms of the PESA is necessary.

\subsubsection{The action of dynamical similarity} 
\label{sub:the_action_of_dynamical_similsubarity}

To understand how dynamical similarities act on the basic structures of a theory, first recall that the action can be written in terms of the Hamiltonian as
\begin{equation}\label{eq:gen S}
	S = \int \de t \lf( \dot q \cdot p - \H  \rt)\,.
\end{equation}
In order for the pullback of $S$ to have the form \eqref{eq:gen dyn sym} under $\mS$, we must have that the phase space quantities below transform as
\begin{align}
	\mS^* p\de q &= c p \de q + \de \phi \\
	\mS^* \H\de t &= c \H \de t + \de \phi\,.
\end{align}

We further restrict $\mS$ to be such that it has a group action on $\Gamma$. Under this restriction, all symmetry transformations of $\mS$ depend only on upon the instantaneous state of the system (i.e., the values of $q$ and $p$), the choice of Hamiltonian $\H$ and the time parameter $t$. Moreover, such a symmetry will map the structures of a theory to structures in the same theory. The group action of $\mS$ on $\Gamma$ can be conveniently expressed in terms of the infinitesimal representations:\footnote{ For simplicity in this paper we work with explicit expressions in terms of Darboux coordinates on $\Gamma$. For the more geometrically inclined reader, these definitions are equivalent to requiring that the tangents, $v_D$, to the orbits of $D$ in $\Gamma$ are defined implicitly as the unique solution to: $\iota_{v_D} \omega = \theta + \de\phi$, where $\omega$ and $\theta$ are the symplectic 2-form and symplectic potential respectively and $\iota$ denotes the interior product on the exterior algebra of $\Gamma$. }
\begin{subequations}\label{eq:local S}
\begin{align}
	\delta_\mS q &= n q + \pb{q}{\phi} & \delta_\mS p &= (1-n) p + \pb p \phi \label{eq:local pq}\\
	\delta_\mS \H &= \Lambda\H + \pb \H \phi & \delta_\mS t &= (1- \Lambda) t \label{eq:local Ht} \,,
\end{align}
\end{subequations}
for the arbitrary real numbers $n$ and $\Lambda$, and where the \emph{Poisson bracket} is defined by $\pb f g = \diby f q \diby g p - \diby g q \diby f p$.\footnote{ Note that the equations \eqref{eq:local S} are partial differential equations for $\phi$ on phase space. We are not aware of any general existence theorems for these equations, but we can show that solutions exist for all the examples treated in this paper. }

The first line \eqref{eq:local pq} indicates that the dynamical similarity $\mS$ rescales the coordinates $q$ and $p$ of phase space in such a way that the combination $q \cdot p$, which has the units of the action (or angular momentum), gets rescaled by a constant. The transformation $D$ therefore relates representations of the system with different conventions for the unit of angular momentum --- in line with our general expectations for a dynamical similarity. The extra Poisson bracket terms involving $\phi$ in the first line reflect the freedom to absorb units differently into either the momentum variables $p$ or the configuration variables $q$. The function $\phi$ encodes all the different possibilities for doing this. For example, $n$ can be redefined arbitrarily to $n\to n + m$ by shifting $\phi$ by $m p\cdot q$.\footnote{ This freedom reflects the ability to redefine a dynamical similarity up to a canonical transformation. In many concrete problems it is useful to make use of this freedom to find the appropriate representation of dynamical similarity for the system in question. }

One important consequence of dynamical similarities being non-symplectic symmetries is that they do not preserve the Liouville measure $\mu_L$. This can easily be seen by noting that the transformations \eqref{eq:local pq} rescale the coordinates $q$ and $p$ in such a way that the integrand of \eqref{eq:Liouville} is necessarily rescaled. This means that the natural notion of volume on phase space is not invariant under dynamical similarity. Some of the consequences of this will be explored in Section~\ref{sub:friction_janus_points_and_attractors} and studied explicitly in the applications of Section~\ref{sec:applications}.

In general the Hamiltonian, $\H(q,p,k_i)$, will be a function both of the phase space variables $(q,p)$ and of certain coupling constants $k_i$, which often carry units themselves. If we assume for simplicity that these couplings are invariant under $\mS$,\footnote{We have already stated that this assumption is not particularly well-motivated but will forgo a more detailed analysis of the more general case for future work.} then $\H$ will transform in a fixed way once a choice for $\phi$ is made in the first line. The second line \eqref{eq:local Ht} should therefore be regarded either as a consistency requirement with the first line or as an independent equation that fixes $\Lambda$ in terms of $\phi$ and $n$. We will see in concrete examples, including the all-important cosmological case of Section~\ref{sub:friedmann_lema^itre_robertson_walker_cosmology}, that satisfying \eqref{eq:local Ht} can lead to highly restrictive conditions on the possible choices of $\phi$, $n$ and $\Lambda$.

\subsubsection{The subsystem context} 
\label{sub:the_sub_system_context}

While most of this section is concerned primary with treating dynamical similarity in the universal context, it can also be useful to understand the considerations that are relevant for dynamical similarity in the subsystem context. In general, if a theory has a subsystem symmetry (recall our definition in Section~\ref{sub:contexts_of_symmetry}) then the Lagrange density of a theory can be decomposed as
\begin{equation}
	\L = \L_\text{sub} + \L_\text{inv} + \L_\text{int}\,.
\end{equation}
 In this decomposition, the term $\L_\text{inv}$ only depends upon the structures that are invariant under the symmetry, the term $\L_\text{sub}$ only depends upon the structures on which the subsystem symmetry acts, and the interaction term $L_\text{int}$ involves the interaction of these two structures.

 The invariance property of $\L_\text{inv}$, the transformation properties of $\L_\text{sub}$ and the linearity of the variations $\delta$ imply
 \begin{equation}
 	\mS^*\delta S \to \mS^*\delta S_\text{inv} + \mS^*\delta S_\text{sub} + \mS^*\delta S_\text{int} = \mS^*\delta S_\text{int}\,,
 \end{equation}
for any value of $c$ and $\phi$. We therefore find that a sufficient condition for a system to have an exact (or approximate) subsystem symmetry is that the interaction term, $S_\text{int}$, between the subsystem structures and the invariant structures be exactly (or approximately) zero. Moreover, this condition does not depend on whether the symmetry is symplectic or not.\footnote{Thus, dynamically similar systems can display \emph{subsystem-recursivity} as defined in \cite{pittphilsci16623}. Note however that a dynamical similarity that acts on a subsystem is not a dynamical similarity of the entire system even if it is a symmetry.}

The universal/subsystem distinction is thus indifferent to the symplectic properties of a symmetry. Intuitions about subsystems gained by studying standard gauge symmetries can therefore remain more-or-less intact when studying dynamical similarity. This can be seen explicitly in the Galilean ship example of Section~\ref{sec:Gal ship} and the Keplerian example of Section~\ref{sub:DS as subsystem}. In both examples, it is the dynamical isolation of the subsystem that permits a subsystem symmetry.


\subsection{Contact geometry and dynamically similar evolution} 
\label{sub:geometric_structure}

In this section we will see how to construct a system of dynamically-similar invariants for a conservative Lagrangian system. The invariant description will no longer be conservative and will display friction-like behaviour. To see this, let us begin with Hamilton's equations, which specify the DPMs for the dynamical system with the action \eqref{eq:gen S} in terms of curves on phase space that satisfy:
\begin{align}\label{eq: conservative ham}
	\dot q &= \frac{\partial \H}{\partial p} = \pb q \H & \dot p &= -\frac{\partial \H}{\partial q} = \pb p \H \,,
\end{align}
where the Hamiltonian $\H$ has no explicit dependence on $t$. It is straightforward to show that the evolution equations preserve the value of the Hamiltonian, which is interpreted as the total energy $E$ of the system, and that the system is therefore conservative.

As we have defined them in Section~\ref{sub:the_action_of_dynamical_similsubarity}, dynamical similarities have a one-dimensional group action on phase space parametrised by the group parameter $c$. We therefore expect $D$ to define one-dimensional orbits on the phase space $\Gamma$, and that the space of dynamically-similar invariants should be odd-dimensional with one dimension less than $\Gamma$. For this reason, \emph{contact geometries}, which are odd-dimensional, are an ideal tool for representing such systems. We will now briefly review these geometries. See the references cited at the beginning of Section~\ref{sec:dynamical_similarity} for more details and proofs.

Contact geometries are constructed to be such that vector fields on them can always be divided into a vertical subspace, which can loosely be thought of as spanning the odd dimension, and an horizontal subspace, which is even dimensional and symplectic.\footnote{ The vertical subspace is simply the kernel of $\de \eta$ (with $\eta$ defined by Equation~\ref{eq:eta darboux}) and the horizontal subspace by the kernel of $\eta$. } According to a theorem by Darboux, a canonical choice of local coordinates $(Q, P, A)$ can always be found such that $(Q,P)$ are canonically conjugate coordinates along the horizontal directions and $A$ is a coordinate along the vertical direction. The geometry of a contact manifold is specified by a \emph{contact 1-form} $\eta$ that can be written in these canonical coordinates as
\begin{equation}\label{eq:eta darboux}
	\eta = P\de Q - \de A
\end{equation}
in analogy to the symplectic potential $\theta = p \de q$ of a phase space. In the same way that the geometry of a symplectic manifold is specified by the symplectic 2-form, the geometry of a contact manifold is specified by the contact 1-form. Moreover, just as the symplectic 2-form is required to be such that the Liouville measure is non-zero, the contact 1-form is required to be such that the volume-form 
\begin{equation}\label{eq:contact vol}
	\mu_c = \eta \wedge (\de \eta)^{N-1}
\end{equation}
on the contact manifold is non-degenerate. In fact, this non-degeneracy requirement is a necessary and sufficient condition for $\eta$ to be a contact 1-form.

The first goal is to find a maximal set of independent phase space functions that are invariant under dynamical similarity. Once this is a achieved, the challenge is to find an autonomous set of evolution equations that will evolve the invariant quantities in the same way as Hamilton's equations \eqref{eq: conservative ham}. In this way, the new system will evolve identically to the original system as far as any invariant quantities are concerned. It is in finding such evolution equations that contact geometry is useful. Towards this end, it will be useful to construct a contact 1-form directly in terms of these invariants.

A simple way to construct the required invariant 1-form is to follow the procedure used in Section~\ref{sub:kepler_s_third_law} and Appendix~\ref{sec:homo V}, which mirrors the general procedure developed in \cite{Sloan:2018lim}. In the examples considered in those sections, we found a function on phase space, which in general we can call $\rho$, that scales in the same way as the Lagrangian under $D$. Then we use this function to rescale all other quantities on $\Gamma$ in a way that the combination is invariant under $D$. For example, an invariant can be constructed from the original Hamiltonian by taking the combination
\begin{equation}\label{eq:contact ham gen}
	\H^c = \frac{\H} { \rho^\Lambda }\,,
\end{equation}
where $\Lambda$ is determined by the choice made in \eqref{eq:local Ht}. Using a bit of ingenuity, or the geometrically inspired procedure of Footnote~\ref{ftn:geo contact}, one can obtain a full set of invariants from the original phase space functions $q$ and $p$ by multiplying by powers of $\rho$ such that $\eta$ takes the standard form \eqref{eq:eta darboux}. The difficulty, of course, is not in finding an invariant set, but one in which a contact form $\eta$ can be constructed that can lead to a simple set of evolution equations.

To construct such evolution equations we are guided by contact geometry. On a contact manifold, if one is given some function $\H^c$, which we will call a \emph{contact Hamiltonian}, then it is always possible to construct a \emph{contact vector field} using $\eta$ that is generated by $\H^c$.\footnote{ More concretely, the invariant contact form can be constructed from some phase space function $\rho$ satisfying $\Lie_D \rho = \rho$ using $\eta = \frac{\iota_D \omega }\rho$. The contact Hamiltonian is related to the projection, $p_H$, of the Hamilton vector field of the Hamiltonian function $v_H$ onto the orthogonal compliment of $D$ via $\H^c = \iota_{p_H} \eta$. See \cite{Bravetti_2017} or \cite{2019JMP....60j2902D} for more details on how to construct the contact Hamiltonian.\label{ftn:geo contact}} This procedure is analogous to how one can use the usual Hamiltonian function to generate Hamilton's equations using the symplectic 2-form in standard Hamiltonian mechanics. Using this procedure, one can pick a contact Hamiltonian $\H^c$ such that the evolutions equations are equivalent to \eqref{eq: conservative ham}. It turns out that a good choice for $\H^c$ is given by \eqref{eq:contact ham gen} and the equations of motion take the form:
\begin{align}\label{eq:contact ham}
   \dot Q &= \frac{\partial \H^c}{\partial P} & \dot P &= -\frac{\partial \H^c}{\partial Q} - P \frac{\partial \H^c}{\partial A} & \dot A &= P \frac{\partial \H^c}{\partial P} - \H^c \,. 
\end{align}
While the explicit form of the equations depends on the choice of $\rho$, it can be shown\footnote{ Since $\rho$ is defined by $\Lie_D \rho = \rho$, any other function $\rho'$ must be related to $\rho$ by $\rho' = \alpha \rho$, where $\alpha$ is an invariant: $\Lie_D \alpha = 0$. The contact form $\eta'$ defined via the procedure in Footnote~\ref{ftn:geo contact} can then be related to $\eta$ by performing a contact transformation. The projection $p_H$, and therefore $\H^c$, is then also invariant up to a contact transformation. } that different choices of $\rho$ lead to an equivalent set of equations of motion. We are thus lead to a unique autonomous system of equations (up to coordinates transformations that preserve the contact form) in terms of a set of invariants of $D$.\footnote{Recently, it has been emphasised in \cite{Sloan:2020taf} that the equations of motion \eqref{eq:contact ham} can be obtained by a variational principle first proposed by Herglotz \citep{herglotz,guenther1996herglotz} in which the action only involves invariants of $D$.} While the system \eqref{eq:contact ham} is explicitly non-symplectic, it can nevertheless be cast as a theory that satisfies evolutions equations of the form \eqref{eq:gen ev}. The PESA is therefore straightforwardly applicable to dynamically similar systems.

\subsection{Friction, Janus points and attractors} 
\label{sub:friction_janus_points_and_attractors}


In this section we study some of the consequences of the invariant evolution equations \eqref{eq:contact ham}. Because the space of invariants is a contact manifold and not a symplectic manifold, the equations of motion \eqref{eq:contact ham} cannot be written in the simple Poisson-bracket form \eqref{eq: conservative ham}. Instead, friction-like terms linear in the momenta $P$ appear in the equations for $\dot P$ and $\dot A$. These arise as advertised because of the non-symplectic character of $D$. Unlike the symplectic equations \eqref{eq: conservative ham}, the friction-like equations \eqref{eq:contact ham} are not invariant under the usual time-reversal operation. We discuss the potential implications of this for addressing the problem of the arrow of time in Section~\ref{sub:aot}.

One way to understand the friction-like behaviour of the system physically is to consider that the contribution of the total energy of the system due to $\rho$ is not included in the invariant system because $\rho$ is used to set the standard of length, and is therefore not itself empirically accessible. Removing $\rho$ from the system therefore removes its contribution to the total energy in the redundant representation. The energy of the remaining system must then compensate for this, and this compensation appears in the friction-like terms of \eqref{eq:contact ham}. In a genuine frictional system, the energy losses due to friction are interpreted as being dissipated into degrees of freedom that are epistemically inaccessible for practical reasons. In contrast, in dynamically similar systems changes in energy can either be losses or gains depending on the dynamics of $\rho$, and these changes are due to the impossibility of being able to observe of the overall scale of the system. We thus use the word `friction-like' to describe the evolution equations \eqref{eq:contact ham} since the analogy to friction holds only at the formal level.

We now investigate some of the consequences of this friction-like behaviour. As already discussed in Section~\ref{sub:the_action_of_dynamical_similsubarity}, the Liouville measure --- i.e., the natural notion of volume on phase space --- is rescaled $\mu_L(R) \to c^N \mu_L(R)$ under $\mS$. This means that the volume of a region $R$ in phase space calculated with the Liouville measure will attribute a volume to states related by $\mS$ as if they were ontologically distinct. In the universal context, the necessity criterion \eqref{crit:necessity} of the PESA for empirically adequate theories implies that, when the theory is empirically adequate, states related by $\mS$ should be treated as identical. The Liouville measure therefore introduces a distinction without a difference and is no longer the appropriate measure to use for assigning volumes. What then are the properties of measures that are compatible with the PESA in the universal context?

To answer this question, we must determine how the evolution of a measure projects onto the invariant space. To be concrete, let us consider the natural volume form $\mu_c(R)$ on a contact manifold defined in \eqref{eq:contact vol} using $\eta$. In the original Hamiltonian system, the Liouville measure is particularly interesting because it is preserved under the evolution equations \eqref{eq: conservative ham}. The dynamically similar measure $\mu_c(R)$, however, is not. To see this, we use the equations \eqref{eq:contact ham} to evolve $\mu_c(R)$. A short calculation (see \cite{Bravetti_2017}) leads to the result:
\begin{equation}\label{eq:focusing}
	\dot \mu_c(R) = - N \diby{\H^c}{A} \mu_c (R)\,.
\end{equation}
Thus, when the frictional term in $\dot P$ of \eqref{eq:contact ham} is not zero --- i.e., when $\diby{\H^c} A \neq 0$ --- then the natural volume form on the invariant space evolves according to \eqref{eq:focusing}. In particular, when  $\diby{\H^c} A > 0$ we have standard friction-like behaviour with the measure shrinking in time. In contrast, when $\diby{\H^c} A < 0$, the measure is growing in time. This shrinking and growing can be understood as a kind of focusing (or defocusing) of the dynamical trajectories on the invariant space. 

The measure-focusing we have just described has important consequences for many kinds of scientific inferences made in a theory. When evaluating a particular piece of evidence --- for instance in the context of a Bayesian update of some prior --- it is necessary to know the likelihood of that evidence given the theory. These likelihoods are probabilities that a particular state is observed given a particular theoretical model of the system, and are therefore represented as (normalisable) measures on the state space of that theory. But in the universal context, the appropriate measures to use are measures on the invariant state space. As we have just seen, such projections can have very different behaviour when compared to measures on the redundant space. These considerations can therefore have a dramatic effect on any inference made within a theory. In cosmology, this becomes particularly relevant as we will explore more fully in Section~\ref{sub:cosmo probs}.

\paragraph*{Janus Points and attractors}

To end this section, we consider two kinds of privileged structures on the state space of a theory described by the friction-like evolution equations \eqref{eq:contact ham}. Before doing so we note that, because of the relation \eqref{eq:focusing}, the quantity $\diby{\H^c}{A}$ at a particular point along a dynamical trajectory gives the rate of expansion (or contraction) of the measure $\mu_c$ in an infinitesimal region around that point. Using this interpretation, we consider points along a dynamical trajectory at which $\diby{\H^c}{A}$, and therefore $\dot \mu_c$ (in an infinitesimal region around that point), changes sign. When this occurs, the canonical measure goes from focusing to defocusing or vice versa. In \cite{Barbour_2014,barbour2013gravitational,barbour:janus}, it is argued that the direction in which the focusing of the measure occurs can be taken to define an arrow of time for observers in systems moving along the trajectories in the focused regions. They call points where such an inflection occurs \emph{Janus} points, in reference to the two-faced Roman god of transitions and time, because different local arrows of time emerge on either side of such points.

Janus points come in two different kinds. The first kind, which we call \emph{type~1}, occurs when $\diby{\H^c}{A} = 0$ but where $\ddot\mu_c \neq 0$. In this case, the measure has an extremum at the Janus point but this extremum is not a fixed point of the measure. These conditions ensure that two different local arrows of time will emerge on either side of the Janus point. An example of a type-1 Janus point is found in the Newtonian $N$-body gravitational system with zero total energy and zero angular momentum treated in \cite{Barbour_2014}.\footnote{ Note that these authors do not distinguish between type-1 and 2 Janus points, which is a language we have introduced in this text. } In these systems a unique type-1 Janus point occurs for all solutions when the moment of inertia of the system is minimised.

The only other way for $\dot \mu_c$ to change sign when moving through a state space point $x$ along a dynamical trajectory is if it runs through infinity at $x$. More precisely, we define a \emph{type-2} Janus point as a state space point $x$ along a dynamical trajectory such that $s \diby{\H^c}{A}(x^-) \to \infty $ and $-s \diby{\H^c}{A}(x^+) \to \infty $, for $s \in \{ -1 , +1 \}$ and where $x^-$ approaches $x$ from below and $x^+$ exits $x$ from above. An important example of a type-2 Janus point can be shown to occur in symmetry-reduced models of general relativity at the big bang. These models are convenient for describing general relativity near a singularity and will be studied explicitly in Section~\ref{sub:sing res in GR}. There, the invariant system will be seen to evolve smoothly at the Janus point, resolving the usual big bang singularity.

A second class of privileged state occurs when there is a region $R$ in the invariant space that is such that the asymptotic dynamical flow of the measure converges: $\lim_{t\to \infty} \dot \mu_c(R)\to 0$. In this case, the region $R$ is a trapped attractive surface or \emph{attractor}.\footnote{ See, for example, \cite{hurley1982attractors} for a technical definition. } When such a region exists, there is a basin of attraction $B(R)$ where states evolve dynamically in such a way that the measure is always shrinking as $R$ is approached. The converse; i.e., a region away from which the measure is always growing; is called a \emph{repellor}. Attractors in particular are important in the study of dynamical systems as they signal the emergence of stable structures that may be used as records (these stable structures may themselves not be static). We thus find that the friction-like behaviour of dynamically similar systems entails a rich set of non-trivial structures --- Janus points, attractors and repellors --- that don't exist in conservative systems. We now study how these structures appear in important applications of general relativity.

\section{Dynamical similarity in the universe} 
\label{sec:applications}

In this section we apply our analysis of symmetry in the context of modern cosmology. We argue that standard practice in cosmology is consistent with the universal context of dynamical similarity. We use the PESA to construct a set of Poincar\'e observables for cosmology and show that these observables are necessary and sufficient for producing an empirical adequate theory of the background cosmic geometry. The PESA then implies that cosmological models that agree on their Poincar\'e observables should represent the same physical state.

We then apply this analysis to a more general class of homogeneous geometries that are believed to approximate the generic behaviour of general relativistic space-times near a singularity. Remarkably, the autonomous evolution of the Poincar\'e observables for these systems is well-behaved at the singularity. This suggests a general procedure for resolving singularities in general relativity using dynamical similarity. We also use the measure focusing in these systems to argue for a new proposal for an explanation of the arrow of time in terms of Janus points and attractors.

\subsection{Conceptual problems in cosmology}\label{sub:cosmo probs}

Cosmological solutions to Einstein's equations are found by applying the so-called \emph{cosmological principle}. This principle roughly states that on sufficiently large scales the properties of the universe are the same for all observers. This implies certain symmetries among the largest structures of the observable universe. Most typically, these symmetries include homogeneity and isotropy. Homogeneity singles out a privileged notion of simultaneity, and this can be used to demarcate standards of length and duration. The intuitions developed in the previous sections can therefore be imported to the relativistic case.

Many conceptual problems in cosmology involve looking for explanations that increase the credence of a certain hypothesis as compared to some `typical' class of counterfactuals. For example, inflation is said to solve the flatness problem because, given the assumptions of inflation, the observed relative flatness of the current universe is, it is claimed, significantly more likely than if inflation had not happened. Other conceptual problems in cosmology, such as the \emph{horizon} or the \emph{cosmological constant} problems, have a similar logical structure but applied to observations regarding correlations between apparently causally disconnected regions of space-time and the value of cosmic acceleration respectively.

What is necessary to formulate all of these problems is a measure on the space of DPMs of the theory. This measure can be used to compute likelihoods within the theory to quantitatively evaluate explanatory claims. In particular, a measure for computing the likelihood of inflation given some matter field dynamics is essential for evaluating any claims about inflation. But as we saw in Section~\ref{sub:friction_janus_points_and_attractors}, dynamical similarity implies that the natural measure is time-dependent. The resulting measure focusing therefore significantly affects any explanatory inference made in cosmology if dynamical similarity is found to be a universal symmetry.

To illustrate these implications, we apply our formalism to homogeneous and isotropic cosmologies and compute the measure focusing in the natural measure of those theories. We find that the measure focusing due to dynamical similarity leads to surprising insights regarding the nature of well-known measure ambiguities for such systems.

\subsubsection{Dynamical similarity in Friedmann--Lema\^itre--Robertson--Walker cosmology} 
\label{sub:friedmann_lema^itre_robertson_walker_cosmology}


Let us now apply our formalism to a theory that can describe the first-order behaviour of our observable universe after the onset of inflation. The theory in question is flat Friedmann--Lema\^itre--Robertson--Walker (FLRW) cosmology. This theory assumes a flat, homogeneous and isotropic spatial geometry coupled to scalar matter fields $\phi_i(t)$ with an arbitrary potential $V(\phi)$. A complete mathematical treatment of dynamical similarity in this system is given in \citet{Sloan:2019wrz}.

An Ansatz for the metric of the flat FLRW theory can be expressed in terms of the volume of (a fiducial cell within) the spatial manifold, $v$:
\begin{equation}
\de s^2 = -\de t^2 + v^{2/3} \left(\de x^2 + \de y^2 + \de z^2\right)\,.
\end{equation}
Readers may be more familiar with this metric expressed in terms of a scale factor $a=v^{1/3}$. We have chosen to use the volume instead of the scale factor in this case as it is more convenient for constructing the invariant algebra and equations of motion.

The Lagrangian for this theory is the Einstein-Hilbert action plus the usual scalar field action $S=\int \sqrt{-g} \lf[R - \frac 1 2 \nabla \phi \cdot \nabla \phi - V(\phi)\rt]$ restricted to homogeneous and isotropic fields. Rendered in terms of our choice of variables this leads (after removing a boundary term) to the Lagrangian density
\begin{equation}\label{eq:FLRW L}
\L = v \left(-\frac{2}{3} \frac{\dot{v}^2}{v^2} + \sum_i \frac{\dot{\phi}^2_i}{2}  - V(\vec{\phi}) \right)\,.
\end{equation}
It is straightforward to find a representation of the dynamical similarity transformation, $D$, for this system as it acts only on the volume and not on $t$. Thus, $D: v \rightarrow \alpha v$, which suggests using the spatial volume $v$ as the standard of length for the system. That $v$ is not an invariant of the original system of variables and can thus be used to set a length standard for the system is somewhat unsurprising, and relates to a well-known feature of flat FLRW cosmology that solutions are independent of an overall choice of scale factor (or volume).

Following the procedure described in Section~\ref{sub:geometric_structure} and using $\rho = v$, we find a simple representation of the  Poincar\'e observables to be the Hubble expansion parameter $H=\frac{\dot{v}}{3v}$, the scalar fields $\phi_i$ and their invariant velocities $\dot \phi_i$.\footnote{ A representative of the symplectic potential for the FLRW theory is $\theta_\text{FLRW} = - v \de H + \pi_\phi^i \de \phi_i$, where $\pi_\phi^i = v \dot \phi^i$. Using $\rho = v$, we find an invariant contact form: $\eta_\text{FLRW} = - \de H + \dot\phi^i \de \phi_i $, which leads to the invariant algebra above. The contact Hamiltonian is then $\H^c_\text{FLRW} = \H_\text{FLRW}/v $ and the contact equations of motion \eqref{eq:contact ham} can be shown to lead to the equations \eqref{Klein-Gordon}. Note that $\H_\text{FLRW}$ vanishes on dynamical trajectories, which leads to some simplifications.\label{ftn:FLRW geom} } Given the invariance of $t$ and $\phi$, the only non-trivial invariant is the Hubble parameter $H$. This encodes relative changes of scale in the universe and is observable in terms of the red-shifting of light emitted from distant galaxies. Thus although the overall scale (expressed in terms of volume or scale factor) cannot be determined, its relative change (i.e. the Hubble parameter) can.

To understand how this relative change is manifested in a dynamically similar terms, we can examine the dynamical equations of the theory. In terms of the Poincar\'e observables listed above, the contact equations \eqref{eq:contact ham} for this system can readily be computed (see Footnote~\ref{ftn:FLRW geom} for more details). They are found to be expressible as the well-known Klein-Gordon and Friedmann equations of standard cosmology:
\begin{equation}
\label{Klein-Gordon} \ddot{\phi_i} + 3H \dot{\phi_i} + \frac{\partial V}{\partial \phi_i}=0 \quad H^2 = \frac{8\pi}{3}\left(\sum_i \frac{\dot{\phi}_i^2}{2} + V(\vec{\phi})\right)\,.
\end{equation}
In \cite{Sloan:2020taf} these were obtained using a variational principle due to Herglotz which makes no reference to the volume $v$. The set of invariants described by the dynamical system above can be expressed as an autonomous system, for example, by using one of the scalar fields as an internal clock. Once expressed in this way, the FLRW theory has been written as a set of autonomous equations invariant under dynamical similarity.

The equations we have obtained are the same equations used to described the evolution of the first order background fields in inflationary cosmology. Given that inflationary theory is empirically adequate and, in particular, that a specification of the value of $v$ is not required in inflationary theory for empirical adequacy, the necessity criterion \eqref{crit:necessity} of the PESA and empirical adequacy suggest that \emph{dynamical similarity is a universal symmetry of inflationary theory}. Put differently, current best practice in cosmology is operating in the universal context of dynamical similarity. In this context, representations of states differing by a dynamical similarity transformation, and therefore by a choice of $v$, should be physically identified. While the gauge character of $v$ is widely acknowledged by cosmologists, many of the consequences of treating dynamical similarity as a universal symmetry discussed in Section~\ref{sub:friction_janus_points_and_attractors} are not. We consider some of these consequences now.

\subsubsection{Measures focusing, Janus points and attractors in FLRW}
\label{sub:measure fosusing in FLRW}

The most important characteristic of the scalar field equations~\eqref{Klein-Gordon} is the friction-like character of the term linear in $\dot\phi_i$. To understand the effects of this term, consider choosing a quadratic potential for $V(\phi)$. In this case, the equations are mathematically identical to the equations of motion of a set of harmonic oscillators subject to a frictional term proportional to $H$. In a representation in terms of $v$, the term linear in $\dot \phi$ is seen as due to the expansion of universe via the (re-scaled) velocity $H$ of $v$. This can be used to model the red-shift phenomena often associated with expansion. In the invariant picture, the red-shift is described in terms of a damping of the frequency of the scalar field oscillations. Universal expansion in this context is therefore accounted for in the scale-free system as a friction-like term in \eqref{Klein-Gordon}. Moreover, it is only through this frictional term that the $\phi_i$ fields interact: were the Hubble parameter zero, there would be no coupling between the fields at all. The frictional term is therefore responsible for much of the non-trivial behaviour of the system.

To understand the effect of this frictional term in FLRW cosmology, let us examine its implication towards the construction of measures. Using our choice of variables, the scalar fields and their momenta are invariants of $D$. Since these form a canonical pair, the only non-symplectic element of the invariant algebra is the Hubble parameter $H$. As such the contact form can be represented as $\eta=-dH + \dot{\phi}\, \de\phi$ (see Footnote~\ref{ftn:FLRW geom} for more details). As discussed in Section~\ref{sub:friction_janus_points_and_attractors}, the natural notion of volume on the invariant space is $\mu_c = \eta \wedge d\eta^{N}$, where $N$ is the number of scalar fields. For simplicity of exposition we will restrict ourselves to the case of a single field.

Our goal is to use $\mu_c$ to compute a measure on the space of DPMs of the theory. To do this, we must first remember that $\mu_c$ is a measure on the invariant state space, and not the invariant space of models. We can construct the space of models by choosing a clock internal to the system and then use this clock to define a surface of initial values. Such a surface will then parametrise the space of DPMs of the theory because each choice of initial state with the same value of the internal clock is an initial condition for a different DPM. We can then define a measure on the space of DPMs by restricting $\mu_c$ to this surface. As a final step, we use the fact that along a DPM the system obeys the Friedman equation (i.e., the second equation of \eqref{Klein-Gordon}). If we use a surface of constant red-shift $H = H_0$ to define an internal clock and solve $\dot\phi$ using the Friedman equation, we can define the following measure on the space of DPMs for the single-field FLRW theory:
\begin{equation}\label{eq:GHS measure}
\Omega_{DS} = \sqrt{\frac{3H^2_0}{4\pi} - V(\phi)} d\phi\,.
\end{equation}

This measure is very similar to that first developed in \cite{Gibbons:1986xk} (GHS), and later used in \cite{Gibbons:2006pa} to argue that inflation is fine-tuned. The key difference is that the GHS measure was applied to the space of DPMs of the original symplectic system and therefore included $v$: $\Omega_{GHS}=\Omega_{DS} \wedge \de v$. The difficulty with this measure is that the range of $v$ is unbounded, and therefore any measure defined in this way is subject to regularisation ambiguities. These regularisation ambiguities can be seen through the $H_0$ dependence of $\Omega_{DS}$: the measure depends on the choice of initial value surface unlike the Liouville measure of symplectic systems. Ironically, the authors note that physical processes such as inflation can be defined without reference to $v$ but do not identify dynamical similarity as an explicit symmetry of the system. More importantly, they do not connect the time dependence of the measure to the non-symplectic character of the gauge symmetry inherent to this system.

When examining the FLRW system as a friction-like system resulting from a non-symplectic symmetry, we can examine the effects of measure focusing discussed in Section~\ref{sub:friction_janus_points_and_attractors}. Applying equation~\eqref{eq:focusing} to this case, we find the simple result:
\begin{equation}\label{eq:vol weight}
\dot \eta = \frac{3H}{4\pi} \eta \rightarrow \eta_2 = \eta_1 Exp \left[ \frac{3}{4\pi} \int_{t_1}^{t_2} H dt \right] \rightarrow \eta_2 = \frac{v_2}{v_1} \eta_1 \,.
\end{equation}
Because the natural dynamically similar measure is constructed from the contact form, dynamical similarity implies a natural `volume weighting' to measures \cite{Sloan:2015bha} (where by `volume' we mean spatial volume of a fiducial cell). Thus, solutions which undergo the most expansion between two surfaces of constant red-shift will be those on which the measure focusses \citep{Sloan:2016nnx}. An immediate consequence of this is that de~Sitter expansion is an attractive fixed point of the dynamics since space begins to expand exponentially near such a point. Moreover, near the big bang at $v \to 0$ the measure diverges and therefore obeys a necessary condition for a type-2 Janus point as defined in Section~\ref{sub:friction_janus_points_and_attractors}. Indeed, in Section~\ref{sub:sing res in GR} below we will see that the evolution of the Poincar\'e observables is well-defined at $v = 0$, and that this does indeed correspond to a type-2 Janus point.

The focusing of the measure when using red-shift as an internal clock is apparent from the $H_0$ dependence of $\Omega_{DS}$ in \eqref{eq:GHS measure}. The ambiguity in choosing a measure on the space of models in FLRW cosmology can then be understood in terms of the time-dependence of the measure due to dynamical similarity. Because there is no preferred instant to define an initial value surface, the time variation of the measure immediately implies a kind of ambiguity. This ambiguity, however, does not extend to the amount of measure-focusing that occurs between two surfaces of constant red-shift, which is fixed by \eqref{eq:vol weight}. Being close to de~Sitter expansion, as is true in the current epoch, thus implies that a significant amount of inflation must have occurred in the past. This lends credence to the hypothesis of an inflationary scenario, as was first pointed out in \cite{Corichi:2013kua}.

Note that the measure-focusing implied by dynamical similarity could have significant implications for many other conceptual problems in cosmology. For example, we will see in Section~\ref{sub:aot} that the time-dependence of the measure can be used to provide an explanation for the arrow of time in terms of the Janus point and attractor of the FLRW theory. Other conceptual problems such as the flatness problem or the horizon problem, which depend upon measures on the space of DPMs in the presence of inhomogeneities, could also be affected. We leave those considerations to future work.

\subsection{Singularity resolution in general relativity}\label{sub:sing res in GR}

Singularities occur in general relativity when the evolution equations become non-prescriptive. The precise mathematical characterisation of a singularity can differ depending on the kind of pathology that one wants to emphasise. In one common usage, a space-time point is said to be singular if some physically significant curvature invariant (e.g., the Ricci scalar) is divergent there. Singularities of this kind, although significant in some regard, are usually not seen as fundamental threats to the theory. A more worrisome notion of singularity occurs when the equations of motion can no longer uniquely determine the evolution of the space-time geometry along time-like geodesics within a finite proper-time. When this occurs, the space-time is called \emph{geodesically incomplete} and general relativity is often said to predict its own demise.

It was proven in the famous Hawking--Penrose theorems of \cite{hawking1970singularities,Hawking:1976ra} that geodesic incompleteness is an ubiquitous feature of cosmological space-times in general relativity. The theorems state that if expansion is occurring and the matter content of the universe obeys certain causally motivated conditions,\footnote{ More specifically, if a finite region along a slicing of the space-time is backwardly trapped and the matter content obeys the null energy condition.} then the space-time will be geodesically incomplete in the past. If the conditions in question are satisfied everywhere on a space-like hyper-surface, then all time-like trajectories that intersect this hyper-surface will have started in a singular region often regarded as a \emph{big bang}.

What is significant about these theorems is that they involve a failure of the continuity of the evolution equations of the system at this singular region. In the cosmological case, they are examples of an application of the  Picard--Lindel\"of theorems,\footnote{ See \cite{lindelof1894application} for the original work and \cite[\S2.9, pp 43-44]{apostol2012modular} for a modern exposition. } which give conditions under which a series of differential equations uniquely determine a solution. Broadly speaking, they show that in a set of (possibly coupled) first order ordinary differential equations, $\dot{x_i} = f_i (x_j)$ there exists a unique solution whenever the functions $f_i$ are Lipschitz continuous in the $x_j$. In the Friedmann equation of FLRW cosmology, for example, this continuity condition is violated as $H \rightarrow \infty$ and $v \rightarrow 0$, and leads to the big bang.

This is where the PESA becomes especially relevant. The PESA demands the essential and sufficient autonomous system required to achieve empirical adequacy. But as we have shown in Section~\ref{sub:measure fosusing in FLRW} above, empirical adequacy can be achieved even when the variable $v$ has been removed. The standard FLRW system does include $v$ and is therefore sufficient but not necessary for producing an empirically adequate theory. One can then immediately ask whether the failure of Lipschitz continuity persists when $v$ is removed from the system? The answer is a resounding NO: the dynamical equations of the invariant system are fully prescriptive through the big bang! Instead, the big bang is replaced by a type-2 Janus point as advertised in Section~\ref{sub:measure fosusing in FLRW}.

This perhaps surprising result was worked out for the FLRW case in \cite{Sloan:2019wrz} and, for more general homogeneous space-times in \cite{Koslowski:2016hds}. To illustrate these results here, we will focus on the algebraically simple case of two free scalar fields evolving on an FLRW background. The simplicity of the example however does not limit the generality of the result. For full details of the more general case, we refer the reader to the referenced papers. 

\subsubsection{Singularity resolution in FLRW space-time}\label{sub:FLRW sing res}

\begin{figure}[h]
\includegraphics[width=\linewidth]{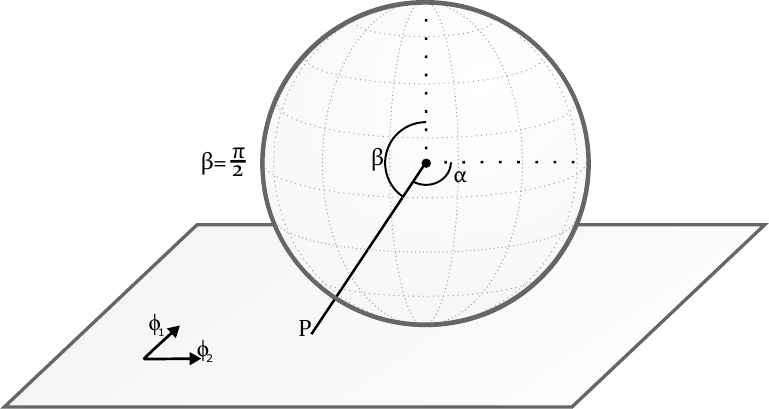}
\caption{The projection from the $\phi_1 - \phi_2$ plane onto the sphere in $\alpha,\beta$. The singularity of general relativity is at $\beta=\pi/2$ which is a regular point of the equations of motion on the sphere.}
\label{Projection}

\end{figure}

To construct an algebra of Poincar\'e observables for the FLRW system, we begin by defining a new set of variables by performing a compactification of the scalar fields to a 2-sphere using the substitution: $\phi_1 = |\tan \beta| \cos \alpha$ and $\phi_2 = |\tan \beta| \sin \alpha$. The singularity of GR lies at $\beta \rightarrow \pi/2$, at which point both $\phi_1$ and $\phi_2$ tend to infinity. The compactification can be seen as taking the $\phi_1 - \phi_2$ plane and placing a sphere tangent to the origin. The point $\phi_1,\phi_2$ is projected onto the sphere by drawing a line from the centre of the sphere to $(\phi_1, \phi_2)$. The variable $\alpha$ corresponds to the longitude on the sphere at which the point intersects the sphere, and $\beta$ the angle this line makes with a line from the origin of the plane to the centre of the sphere as shown in Figure~\ref{Projection}.

One can then examine the equation of motion for $\alpha$ in terms of $\beta$. We note here that $\dot{\phi}_1$ and $\dot{\phi}_2$ also tend to infinity at the singularity. We could consider compactifying them in the same way as we did the positions $\phi_1,\phi_2$, onto a second sphere $(\theta,\chi)$. However, were we to do so, we would quickly see that at the singularity, $\theta \rightarrow \alpha$. By examining the equations of motion for the scalar fields, we see that although $\theta-\alpha$ tends to zero, $\lambda = \frac{\sin(\alpha-\theta)}{\cos \beta}$ remains non-zero and finite. This leads us to express the dynamics of our system in terms of $\alpha$ and $\lambda$:
\begin{equation}
\frac{\de \alpha}{\de \beta} = \frac{\lambda}{\sin \beta \sqrt{1-\lambda^2 \cos^2 \beta}} \quad \frac{\de \lambda}{\de \beta} = \frac{\lambda}{\cos \beta} \left(\sin \beta - \frac{1}{\sin \beta} \right)\,.
\end{equation}
Remarkably these equations remain finite as we approach the singularity; in the limit as $\beta \rightarrow \pi/2$, we see $\frac{\de \alpha}{\de \beta} \rightarrow \lambda$ and $\frac{\de \lambda}{\de \beta} \rightarrow 0$. Because of this, they satisfy the conditions of the Picard--Lindel\"of theorem, and thus have a unique solution about this point. In fact the system of equations can be explicitly integrated. The solutions are geodesics on the sphere. This is the expected geodesic motion on the $(\phi_1,\phi_2)$ plane in the absence of potentials. Since this autonomous system remains well-posed at the point corresponding to the singularity of general relativity, we can use it to continue the solutions beyond this point and thus achieved evolution through the big bang. This evolution, however, goes through a type-2 Janus point as the measure has a divergence at $H\to \infty$ (i.e., at the point corresponding to the big bang in the redundant representation) according to \eqref{eq:vol weight}. Thus, the big bang is replaced by a smooth type-2 Janus point in the dynamical system prescribed by the PESA and the requirement of empirical adequacy.

\subsubsection{General singularity resolution}

An obvious question regarding the results of the previous section is whether they can be generalised to more generic space-times. After all, the power of the Hawking--Penrose singularity theorems is that they apply to generic classes of physically plausible solutions of the Einstein equations. However, a conjecture by Belinkskii, Khalatnikov and Lifschitz\footnote{ See \cite{Belinsky:1970ew,Uggla:2003fp,Ashtekar:2011ck,Belinski:2017fas}.} suggests that the generic behaviour of a physically plausible class of solutions of general relativity near singularities can be described by homogeneous but anisotropic space-times. Such space-times go by the name of \emph{Bianchi models} in which space-time is split into the product of a three-dimensional homogeneous spatial manifold and a one dimensional temporal manifold valued on the real line. The Bianchi models are classified into eleven types (nine labelled by Roman numerals with two subtypes) according to the structure of their spatial symmetries. One can therefore dramatically extend the generality of our results by applying our formalism to the Bianchi models. This has been achieved in \cite{Koslowski:2016hds}, where a rational reconstruction of the analysis suggests that the PESA can be used to produce a singularity-free evolution in terms of a dynamically similar algebra of Poincar\'e observables. We review these results now.

It is convenient to express the Bianchi space-times in terms of \emph{Misner variables}, which are covariant vector fields $\sigma_i$ on the spatial manifold. In terms of these variables, the metric is
\begin{equation}
\de s^2 = -\de t^2 + v^{2/3} \left( \exp \left(-\frac{q_1}{\sqrt{6}} - \frac{q_2}{2} \right) \sigma_1^2+ \exp \left(-\frac{q_1}{\sqrt{6}} + \frac{q_2}{2} \right) \sigma_1^2 + \exp \left(-\frac{\sqrt{2}}{3} q_2 \right) \sigma_3^2 \right)\,,
\end{equation}
where the normalizations of $q_1$ and $q_2$ have been chosen so that when expressed as a Lagrangian theory, their conjugate momenta appear as scalar fields subjected to a potential. This potential is determined by the symmetries of the spatial manifold, but it always takes a standard form as a function of the Misner variables multiplied by the volume raised to the one-third power; i.e., $v^{\frac{1}{3}} V_s (q_1, q_2)$. This homogeneity in $v$ is central in establishing the dynamical similarity of these models because it allows re-scalings of the volume to be understood independently of the remaining variables. Up to a boundary term which removes the second derivatives of $v$ from the equations, the Lagrangian for these models can be written as
\begin{equation}
\L = v \left(-\frac{2}{3}\frac{\dot{v}^2}{v^2} + \frac{\dot{q_1}^2}{2} + \frac{\dot{q_2}^2}{2} - v^{\frac{1}{3}} V_s (q_1, q_2) \right) \,.
\end{equation}
The Lagrangian is very similar to that for the FLRW models \eqref{eq:FLRW L}, though instead of matter fields we have anisotropies, and the coupling to their potential has a different dependence on the volume.

Using the volume scaling of these fields, we can find a representation for the dynamical similarity transformation for this model as $D: t \rightarrow \alpha t, v \rightarrow \alpha^3 v$. This has an intuitive geometrical interpretation: it rescales each of the four dimensions of space-time equally $\de s^2 \rightarrow \alpha^2 \de s^2$.

From the form of $D$ we can find the autonomous algebra, $\mA$, of Poincar\'e observables following the general procedure outlined in Section~\ref{sub:geometric_structure}. One such representation that is also dynamically closed is given by (using $\rho = 2 v^{2/3}$)
\begin{equation}
\mA = \lf\{q_1, \, q_2,\, \Psi_1= v^{\frac{1}{3}}\dot{q}_1,\, \Psi_2= v^{\frac{1}{3}}\dot{q}_2, \, \Phi = -\frac{2\dot{v}}{v^{\frac{2}{3}}} \rt\}\,.
\end{equation}
The equations of motion \eqref{eq:contact ham} take the form
\begin{equation}
\dot{q_i} = v^{-\frac{1}{3}} \Psi_i \quad  \dot{\Psi_i} = v^{-\frac{1}{3}} \left(\frac{\partial V_s}{\partial q_i} - \frac{\Phi \Psi_i}{3} \right) \quad \dot{\Phi} = v^{-\frac{1}{3}} \left(\frac{\Psi_1^2}{2} + \frac{\Psi_2^2}{2} + \frac{\Phi^2}{6} \right)\,.
\end{equation}

As in our previous examples, we see that an autonomous system independent of $v$ (which we've chosen to set the length standard for this system) can be constructed by using any of the above invariants as internal clocks. This is of particular importance in addressing the issue of singularities. The equation of motions for the system, when written in the original set of variables, has terms that are proportional to $v^{-1}$. Hence the dynamical equations are not Lipschitz continuous at $v=0$, and the evolution becomes singular. 

Rather remarkably, in the autonomous system, this problem is bypassed. The invariant equations remain Lipschitz continuous throughout the evolution as they are independent of the problematic variable $v$ and hence can be evolved beyond the apparent singularity. This property has been shown to hold in several systems \citep{Koslowski:2016hds}. In Bianchi systems we can compactify $\Psi_1, \Psi_2$ on a sphere such that $\Psi_1 = |\tan \beta| \cos \alpha$ and $\Psi_2 = |\tan \beta| \sin \alpha$ following the procedure used for the FLRW theory of Section~\ref{sub:FLRW sing res}. Doing so we see that the singularity of GR is at $\beta = \pi/2$. Incredibly, this turns out to be a regular point of the equations of motion of the invariant system.

\subsubsection{Singularity resolution in black holes?}

A natural question to ask at this point is whether the elimination of singular behaviour seen in the previous sections also occurs in the case of black holes. While we will not claim here to provide a definite answer to this question, we will highlight some on-going investigations using idealised models that suggest that dynamical similarity may have important implications for the physics of black holes.

The interior of a black hole corresponds closely to a Kantowski--Sachs cosmology, which has the spatial topology of a 2-sphere crossed with a real direction. Under interchange of the time and radial directions this correspondence is made exact, and so we can examine the evolution of the interior of a black hole towards the singularity by examining the evolution of a Kantowski--Sachs space-time towards its big bang. The Kantowski--Sachs space-times can be represented in a way very similar to the Bianchi models presented above. Applying similar machinery, one can find that an autonomous system invariant under dynamical similarity can be constructed and has a well-defined evolution through the point in the dynamics that is singular in general relativity \citep{sloan-mercati:BHres}.

As this construction produces a well-defined dynamical system when the general relativistic dynamics becomes singular, the question arises as to how this result can be squared with the singularity theorems of Hawking and Penrose. The key to understanding this is to note that the essential and sufficient autonomous system that is well-defined through the general relativistic singularity is not itself well-represented by a Lorentzian space-time. At the point of the singularity, the spatial manifold is found to change its handedness.\footnote{ See \cite{Koslowski:2016hds} for details. } This means that the sign of the determinant of the spatial metric switches sign precisely when the system passes the singular point.

This is fatal for a description of the system in terms of space-time. The geodesic equation requires the determinant of the metric to have a well-defined sign, and therefore breaks down at this point. When this happens, the space-time representation of the system becomes geodesically incomplete in accordance with the Hawking--Penrose singularity theorem. However, in the universal context, a change in the handedness of space leads to an indiscernible change in the system since the internal relations between events remain unchanged.\footnote{ For a lengthy discussion of this point in relation to a classic discussion initiated by Kant, see \cite{pooley2003handedness}. } Geodesic completeness of this kind in therefore not sufficient to ensure that the empirically relevant observables of the system are badly behaved. Indeed, the Poincar\'e observables for this system do have well-defined evolution through the point at which parity is changed, and this is all that is required to describe the physical properties of this system. Of course, it remains to be seen whether such a description of the system can ever be found to be empirically adequate. There are many open questions left to resolve. Nevertheless, the Hawking--Penrose singularity theorems are ineffective at signalling a breakdown of determinism in the dynamically similar theory.

\subsection{The problem of the arrow of time}\label{sub:aot}

At a macroscopic level, processes occur in a way that defines a significant gradient in time that is persistent across the observable universe. No supernova has been seen to implode into a star while a hot cup of coffee cools on a desk. Astrophysical processes and earthly thermodynamic systems alike seem to unfold in a consistently time-asymmetric way. This monotonic gradient in time defines a persistent \emph{arrow of time} across the known universe. Certain quantitative properties of a system --- most commonly the entropy --- can be used to give quantitative meaning to this gradient, and therefore can be used to define time's arrow. More generally, \cite{price2002boltzmann} defines an arrow of time in terms of the following criteria:
\begin{quote}
	``(i) ... an asymmetry of physical processes in time, not with an asymmetry in time itself; (ii) that the objective asymmetry concerned comprises a monotonic \emph{gradient}, rather than an increase or decrease; (iii) that the asymmetry in nature is a matter of numerical imbalance between temporal mirror images, not of literal reversibility; and (iv) that if need be the term `entropy' itself is to be thought of as a kind of variable place-holder for the relevant properties of a vast list of actual physical asymmetries.''[Original emphasis.]
\end{quote}
A problem occurs when attempting to account for the observed arrows of time in terms of the microscopic laws of physics. These laws do not exhibit a significant amount of time asymmetry --- certainly nothing like the amount required to explain the enormous temporal gradients observed across our universe.

A prominent approach to this problem is to postulate a \emph{Past Hypothesis} \citep{albert2009time,goldstein2001boltzmann} whereby the earliest known states of the universe are postulated to be highly atypical. According to the arguments of the Past Hypothesis, if the past states are highly atypical, then the present state is highly likely be vastly more typical because of a tendency of many dynamical systems to evolve towards more typical states.\footnote{See \cite{frigg2009typicality} for a critique of the dynamical aspects of this argument.} Thus, the Past Hypothesis is said to provide an explanation for the existence of a significant gradient in time, which in this case is quantified by the typicality of the state of the early universe.\footnote{ This notion typicality is usually expressed in terms of some version of the Boltzmann entropy of the system.} The Past Hypothesis is thus said to provide an explanation for the arrow of time in terms of fundamentally time-symmetric laws.

What is immediately apparent from the line of reasoning outlined above is that the Past Hypothesis relies very strongly on having an unambiguous measure, stable in time, that can be used to determine the typicality of a particular state of the world. Ambiguity in the measure is problematic because different choices of measure could disagree on the significance or direction of the gradient. Time dependence in the measure is problematic because a time-varying measure could itself be the source of time-asymmetry, and thus undermine the explanatory force of the Past Hypothesis.\footnote{ See \cite{gryb2020new} for more details on about this argument as well as \cite{Curiel:2015oea} and \cite{earman2006past} on ambiguity issues regarding the measure in the Past Hypothesis. }

The arguments of Section~\ref{sub:friction_janus_points_and_attractors}, however, suggest that the natural measures of cosmology, and indeed of any system where dynamical similarity is a universal symmetry, will evolve in time. Moreover, because of Equation~\eqref{eq:vol weight}, the amount of measure focusing between two instants is proportional to the amount of expansion that takes place in the universe between those two instants. This leads to an enormous amount of measure focusing ($\sim 85$ orders of magnitude) between the present epoch and the onset of inflation. Thus, when dynamical similarity is taken into account, the dynamics of the measure itself already provides a gradient in time that is big enough to account for a significant numerical imbalance, and thereby define a cosmological arrow of time according to Price's definition above. Because this arrow of time results from the dynamical equations of the invariant theory, we have a dynamical explanation of a cosmological arrow that does not involve the use of a Past Hypothesis. The dynamical nature of this explanation can be seen from the fact that the friction-like terms of Equations~\ref{eq:contact ham} are not invariant under the time reversal operation $t \to -t$.

It's important to note however that the mere existence of friction-like terms in the measure dynamics is not sufficient to provide an adequate explanation of the arrow of time. One might worry that some form of time-symmetry is encoded in the fact that for every DPM in cosmology there is a matching DPM that is representationally its time-reversed image. Moreover, as we already noted, the significant amount of time-asymmetry observed in our universe must be accounted for. These two issues can be addressed by a scenario first proposed in \cite{Barbour_2014} and expanded upon in \cite{barbour:janus}. We will give a slightly modified account here.

In our account, the space of DPMs of a cosmological theory should contain a dense set of DPMs characterised by a unique Janus point that is in the attractive basin of some attractor for the system. In such a scenario, an observer described by a state in a DPM near the attractor will see an arrow of time pointing from the Janus point to the attractor. The divergent amount of measure-focusing that occurs near the attractor could thus explain the enormous time-asymmetry seen by such observers, and therefore provide the basis for a cosmological arrow of time of the form required by Price.

Moreover, a symmetry of the space of DPMs about the Janus points can then explain how such arrows of time can nevertheless emerge when the system is embedded into a time-symmetric symplectic system. Because the arrows of time are assigned to observes near attractors, two separate arrows of time can then emerge for observers along dynamical trajectories pointing towards different attractors on opposite sides of a Janus point. As we have seen, such a scenario occurs in the cosmological models studied in Section~\ref{sub:cosmo probs} where the generic solutions were seen to exhibit type-2 Janus points in the basins of de~Sitter attractors.

Because Janus points are defined by a dynamical condition on the state space, they need only lie in the attractive basin of the relevant attractor, and can thus belong to a dense set of trajectories which may well be typical according to some choice of measure. Thus, the existence of a Janus point is not a Past Hypothesis itself because Janus points can (and do in the examples explored in \cite{Barbour_2014}) belong to a typical set of trajectories.

An open question remains to show that the cosmological arrow of time defined by measure focusing can be used to recover the local arrows of time (e.g., the thermodynamic and perceptual arrows, etc) familiar to our everyday experience of the world. This is the subject of on-going investigations.

\section{Conclusions} 
\label{sec:conclusions}

\subsection{Dynamical similarity and cosmology} 
\label{sub:dynamical_similarity_and_cosmology}


We have studied several different aspects of dynamical similarity and have shown that there are good reasons to believe that this symmetry has important implications for cosmology. We found that removing the ``extra mathematical hooks'' on which similarity hangs reveals a well-posed system with a smooth Janus point replacing a violent big bang. Attractors resulting from friction-like equations lead to an arrow of time for observers, like us, nearing such attractors. When the universe is stripped of its dynamically similar structure, it appears simultaneously less pathological and more comprehensible.

But are we justified in treating as surplus what seems like vital representational structure in some of our most successful theories? We believe so. To address this question, we have developed a conceptual framework for identifying surplus structure in a theory. The principle we proposed --- the PESA --- prescribes that the observable structure of a theory be the structure that must be specified in the form of input data in order to uniquely solve the evolution equations. When applied to cosmology, the PESA combined with the empirical adequacy of the standard model of cosmology implies that dynamically similar structure is indeed surplus.

The implications for treating dynamical similar structure as surplus are profound and, perhaps, counter-intuitive. It was thus helpful for us to build intuition for dynamical similarity by studying explicit examples. Perhaps the oldest known example of dynamical similarity is illustrated in the Kepler problem, which was studied in Section~\ref{sub:kepler_s_third_law}. What we saw there was that it is important to distinguish between situations in which a symmetry should be seen to relate empirically discernible and indiscernible states. Here too, examples were illustrative. By studying the Galilean boost symmetries in different scenarios, we were able to see that empirically discernible states are related by symmetries only when such symmetries act uniformly on subsystems while leaving the remaining system invariant. This intuition was then applied to the Kepler problem. In every example studied, the PESA was found to reproduce orthodox views of symmetry when the theory under study was found to be empirically adequate.

This, we believe, gives us confidence that the PESA is a good guide for determining surplus structure in a theory. Our main motivation for introducing the PESA, however, was to apply it to dynamical similarity in the cosmological setting, where we've obtained novel results regarding the interpretation of cosmological solutions. The crucial element of the PESA in regards to cosmology is the criterion of necessity. This criterion can often be overlooked because a theory can successfully describe the phenomena even if it has more structure than is strictly necessary. But to describe cosmological phenomena, our analysis shows that dynamically similar structure is simply not necessary. No external scale is required to evolve the system: the evolution of the Poincar\'e observables we defined in Section~\ref{sub:friedmann_lema^itre_robertson_walker_cosmology} exactly match their evolution in general relativity except at the big bang. There, it is only surplus structure that ruins the well-posedness of the evolution equations. When this happens, the ``mental structures that we attach to the phenomena'' \citep{vanFraassen:sym_and_surplus} have the potential to mislead. Stripping ourselves of dynamically similar structure removes the initial singularity and suggests an alternative explanation for a cosmological arrow of time. In cosmology, we may therefore have been mislead to believe in a violent beginning and time-symmetric laws.

\subsection{Further implications of dynamical similarity} 
\label{sub:further_implications_of_dynamical_similarity}

As we saw in Section~\ref{sec:dynamical_similarity} dynamical similarity is a very general symmetry of any Lagrangian system in both the universal and subsystem contexts. This suggests that dynamical similarity may have many more applications to physical systems beyond the limited number of systems studied in this paper. When the system is conservative, we saw that the mathematics of dynamical similarity is generically described by contact, rather than symplectic, geometry. But while contact geometry is less familiar to most physicists than symplectic geometry, contact geometry is nevertheless known to have a rich mathematical structure.

One potential application of dynamical similarity is in the foundation of thermodynamics and the identification of certain thermodynamical quantities to apparently conservative systems. Recently, contact systems with equations of motion of the form \eqref{eq:contact ham} have been shown to display strong formal analogies to thermodynamic systems \citep{Bravetti:2015kpo,Bravetti:2018rts}. Natural geometric structures on contact manifolds can be used to define thermodynamic quantities such as equilibrium states and notions of entropy by taking advantage of measure focusing and attractors. Using these structures, it is then possible to describe thermodynamic processes such as approach-to-equilibrium and entropy-production. In the universal context, dynamical similarity could then be used to write the evolution equations of apparently conservative systems in terms of quantities that have formal analogies to thermodynamic quantities like entropy. Such an analysis may be extremely valuable in studying gravitational systems --- particularly in cosmology. More generally, dynamical similarity and contact geometry may prove to be a valuable way to gain a deeper understanding of the known thermodynamic properties of general relativistic systems such as black holes.

A final important consideration is the implications of dynamical similarity to quantum mechanics. Because dynamical similarities alter the symplectic structure by rescaling the unit of angular momentum, dynamical similarities will have important implications for the commutation relations of quantum mechanics. In the universal context, Planck's constant $\hbar$ must be determined relative to other standards of angular momentum within the system. At the representational level, this should imply a dynamical similarity. It is unclear what the precise consequences of this would be in a universal quantum system, but it is clear that more research into such questions would be extremely valuable. At the classical level, dynamically similar systems display friction-like dissipative behaviour. At the quantum level, this may generalise to non-unitary evolution equations, perhaps in the form of a Lindblad master equation. The implications of this to the foundations of quantum mechanics, particularly with regard to the role of decoherence, should be further investigated.



\section*{Acknowledgements} 
\label{sec:acknowledgements}

We would like to acknowledge many stimulating interactions with Julian Barbour, Flavio Mercati and Tim Koslowski regarding the role of dynamical similarity in the universe and Alessandro Bravetti and Connor Jackman regarding the mathematical aspects of dynamical similarity. We'd also like to acknowledge discussions with Simon Friederich about the arrow of time as well as Karim Th\'ebault and Henrique Gomes about the nature of symmetry. SG's research was funded by a Young Academy Groningen Scholarship.


\appendix

\section{Dynamical Similarity with Homogeneous Potentials} \label{sec:homo V}

As a way to illustrate the generality of the procedure used in Section~\ref{sub:kepler_s_third_law} and to further demonstrate how to apply to the general method described in Section~\ref{sub:geometric_structure}, let us consider the case of particles interacting under a single homogeneous potential.

In this situation, the most general Galilean invariant potential between any two particles is a power of their separation: $V_{12}=|\vec{r}_1 -\vec{r}_2|^n$. To best demonstrate dynamical similarity in this system, will work in a coordinate system in which there is zero net total linear momentum, and the coordinate origin is set to the centre of mass of the system. We will also work in coordinates in which the overall size of the system is determined by only one of our dynamical variables, which we call $r$. To do this, we describe the motion in the $3p-3$-dimensional space, $\mathbb{R}^{3p-3}$, as the product of the positive real numbers, $\mathbb{R}_+$, with coordinate $r$, and the $(3p-4)$-sphere, $S^{3p-4}$, with coordinates $\vec{\theta}$. Let us define $K_s$ to be the kinetic energy of a particle moving on a unit $S^{3p-4}$. For example, on the 2-sphere this is
\begin{equation}
K_s = \frac{1}{2}\left(\frac{\de \theta_1}{\de t}\right)^2 + \frac{\sin^2(\theta_1)}{2} \left(\frac{\de \theta_2}{\de t}\right)^2\,.
\end{equation} 

It is important to note that $K_s$ is independent of $r$ and depends on $t$ only through terms such as $\left(\frac{\de \theta}{\de t}\right)^2$. The total potential can be decomposed into a product of a term dependent only on the radial coordinate, $r$, and one dependent only on the angles, $\vec{\theta}$: $V=r^n V_s(\vec{\theta})$. In the literature it is common to refer to $V_s$ as the \emph{shape potential}. The Lagrangian for our system is rendered 
\begin{equation}
\L =\frac{1}{2} \left(\frac{\de r}{\de t}\right)^2 +r^2 K_s + r^n V_s (\vec{\theta})\,.
\end{equation}

We are now in position to repeat the analysis performed in Section~\ref{sub:kepler_s_third_law}. Under the rescaling $D: r \rightarrow \lambda r$, $D: t \rightarrow \lambda^{\frac{2-n}{n}} t$ we see that $D \L \rightarrow \lambda^n \L$, and will again preserve the stationarity condition of the action. The algebra $\mathcal{A}$ consisting of the invariants of $D$ is readily established. We note that under the action of $D$, each of the angles $\theta_i$ is invariant. Using these invariants and the transformation properties of $t$ and $r$, we can construct the invariant velocities $a=r^{-n/2} \dot{r} = \frac{2}{2-n} \frac{\de}{\de t} (r^{\frac{2-n}{2}})$ and $b_i=r^\frac{2-n}{2} \dot{\theta_i}$ (for $n \neq 2$, we will discuss the $n=2$ case immediately below).  In the case in which $n=-1$ this reduces exactly to the Kepler problem as discussed in Section~\ref{sub:kepler_s_third_law}. 

When $n=2$, we reproduce a Hooke potential. In this case, $t$ is unaffected by the transformation $D$, and thus the angles as well as their velocities with respect to $t$ are automatically members of the algebra of invariants, with the remaining basis member being $\frac{\dot{r}}{r}=\frac{\de }{\de t} (\log r)$. Hence we immediately recover the well-known result that in a system of particles connected by springs (e.g. harmonic oscillators) the period of oscillation is independent of the overall amplitude of oscillation. 

\bibliographystyle{apacite}   
\bibliography{SD}



\end{document}